\documentclass[twocolumn,prd,showpacs,10pt]{revtex4}
\usepackage{graphicx}
\usepackage{dcolumn}
\usepackage{bm}

\newcommand{\square}{\kern1pt\vbox{\hrule height 1.2pt\hbox{\vrule
width 1.2pt\hskip 3pt
\vbox{\vskip 6pt}\hskip 3pt\vrule width 0.6pt}\hrule
height 0.6pt}\kern1pt}
\newcommand{\be}{\begin{equation}}
\newcommand{\beqn}{\begin{eqnarray}}
\newcommand{\ee}{\end{equation}}
\newcommand{\eeqn}{\end{eqnarray}}

\begin{document}

\title{Cosmological Dynamics of a Dirac-Born-Infeld field }
\author{Edmund J. Copeland\footnote{ed.copeland@nottingham.ac.uk}, 
Shuntaro Mizuno\footnote{shuntaro.mizuno@nottingham.ac.uk}, and 
Maryam Shaeri\footnote{ppxms1@nottingham.ac.uk }}
\affiliation{\vspace{.5cm}\\ School of Physics and Astronomy, University of Nottingham, University Park, Nottingham NG7 2RD, UK}

\date{\today}

\begin{abstract}
We analyze the dynamics of a Dirac-Born-Infeld (DBI) 
field in a cosmological set-up which includes a perfect fluid. Introducing convenient 
dynamical variables, we show the evolution equations form an autonomous system 
when the potential and the brane tension of the DBI field 
are arbitrary power-law or exponential functions of the DBI field.
In particular we find scaling solutions can exist when 
powers of the field in the potential and warp-factor satisfy specific relations. A new class of fixed-point solutions are obtained corresponding to points which initially appear singular in the evolution equations, but on closer inspection are actually well defined. In all cases, we perform a 
phase-space analysis and obtain the late-time 
attractor structure of the system. 
Of particular note 
when considering cosmological 
perturbations in DBI inflation is a fixed-point solution 
where the Lorentz factor is a finite large
constant and the equation of state parameter of the DBI field is
$w=-1$. Since in this case the speed of sound $c_s$ becomes constant,
the solution can be thought to serve as a good background
to perturb about. 
\end{abstract}

\vskip 1pc \pacs{pacs: 98.80.Cq}
\maketitle

%

\section{Introduction}

The inflationary paradigm remains to date the most successful explanation for the origin of the observed 
temperature fluctuations of the 
cosmic microwave background (CMB) 
(see, e.g., \cite{Linde,Lyth:1998xn} for reviews).
However, establishing its origin in fundamental theory has not been quite as successful and so for many it remains a fascinating paradigm in search of an underlying theory. 
The favourite candidate for this is String theory and for a nice review of the construction of inflation models in string theory, see \cite{Baumann:2009ni}. 

One interesting  model, recently proposed from string theory is Dirac-Born-Infeld (DBI) inflation 
\cite{Silverstein:2003hf,Alishahiha:2004eh,Chen:2004gc,Chen:2005ad,Shandera:2006ax},
where inflation is driven by the motion of a D3-brane
in a warped throat region of a compact internal space.
In this model, since the inflaton is the position of a D-brane  with a DBI action, its kinetic term is inevitably non-canonical. In addition to this kinetic term,  its effective action includes 
a potential arising from the quantum interaction between D-branes, with  
the  brane tension encoding geometrical information
about the throat region of the compact space.  Because of these novel ingredients, the predictions 
of DBI inflation are quite different from the ones from the standard slow-roll
inflation models and it has led to an intense period of research into the scenario, including work into the background dynamics and linear perturbations \cite{Spalinski:2007dv,Spalinski:2007qy,Chimento:2007es,Ward:2007gs,Spalinski:2007un,Kinney:2007ag,Tzirakis:2008qy,Czuchry:2008km}.

From the phenomenological viewpoint, one of the main reasons why this model has attracted attention is because of its sizable equilateral type of primordial non-Gaussinaity, first pointed by 
\cite{Silverstein:2003hf} and explored further in \cite{Alishahiha:2004eh,Chen:2004gc,Chen:2005ad,Chen:2005fe,Chen:2006nt,Huang:2006eh,Arroja:2008ga,Langlois:2008wt,Langlois:2008qf,Arroja:2008yy,Langlois:2009ej,Gao:2009gd,Chen:2009bc,Us,Mizuno:2009cv,Gao:2009at,Mizuno:2009mv,RenauxPetel:2009sj,Chen:2009zp,Koyama:2010xj,Chen:2010xk}.  Such a large signal can not be obtained in standard slow-roll inflation, hence it opens up the possibility of distinguishing this model from that of 
the standard slow-roll inflation-- although not necessarily distinguishing it from more general slow-roll models going beyond single field inflation as they may also lead to large signals.
Furthermore, because of the dependence of the effective four dimensional string tension on the nature of the compact internal space, it is possible to place  
further  constraints on the parameters related with compactifications such as 
flux numbers. For the most up to date observational constraints on and consequences of DBI inflation, see
\cite{Kecskemeti:2006cg,Lidsey:2006ia,Baumann:2006cd,Bean:2007hc,Lidsey:2007gq,Peiris:2007gz,Kobayashi:2007hm,Gmeiner:2007uw,Lorenz:2007ze,Bean:2007eh,Bird:2009pq,Bessada:2009pe,Fuzfa:2005qn,Fuzfa:2006pn}.
Recently the idea of low scale inflation arising from the DBI action has been invoked to explain the late time acceleration we associate with dark energy
\cite{Martin:2008xw,Ahn:2009hu,Chiba:2009nh,Ahn:2009xd}.

Given the results mentioned above and inspired by the possibility of inflation being an attractor solution in DBI models, we believe there is a need to understand the late-time attractor structure for as general a DBI set-up as possible. It is well known that for a canonical scalar field with a potential, scaling solutions can exist where  the ratio  of the kinetic and potential energy of the scalar field maintain the same ratio, and understanding the stability of these solutions is important in determining the nature of the late-time solutions
\cite{Lucchin:1984yf,Ratra_Peebles,trac,Ferreira:1997hj,Copeland:1997et,vdHCW,Heard:2002dr,Padmanabhan:2002cp,Tsujikawa:2004dp,Calcagni:2004wu,Copeland:2004qe,Copeland:2009be}.
Among the earlier work, the phase-space analysis proposed by \cite{Copeland:1997et}
is particularly powerful because it allows us to make use of suitable dimensionless
dynamical variables, in order to establish the global stability of such scaling solutions. 
For related work which analyzes the dynamics in DBI models with general
inflationary potentials, see \cite{Meng:2004ap,Underwood:2008dh,Franche:2009gk,Franche:2010yj}.

Now, this method for analysing the stability of the solutions has only been applied to the case where the DBI field has a quadratic potential and the associated D-brane is in the anti-de Sitter throat
\cite{Guo:2008sz}. The late-time attractor structure for the case without a potential, known as tachyon cosmology, has been studied extensively, \cite{Aguirregabiria:2004xd,Piazza:2004df,Copeland:2004hq,Gumjudpai:2006hg,Tsujikawa:2006mw,Gong:2006sp,Sen:2008yt,Quiros:2009mz,Li:2010eu}
and the case with another degree of freedom,  has been studied in \cite{Gumjudpai:2009uy,Saridakis:2009uk}.

In this paper, in order to make the cosmological application of scaling solutions in DBI models more complete and as a natural extension of  \cite{Guo:2008sz}, we obtain the late-time attractor structure 
of the system including a perfect fluid plus a DBI field whose potential and brane tension
are arbitrary power-law or exponential functions of the DBI field.

The rest of the paper is arranged as follows. In section \ref{basic-eqn} we present the model and basic equations.
Then, in section \ref{section_powerlaw} we consider the models where
the potential and brane tension 
are arbitrary power-law functions of the DBI field.  Special emphasis is given to a new set of fixed-point solutions which at first site appear singular in the equations of motion, but on closer inspection are well defined. This is followed in section \ref{section_exp} with an analysis of the models 
where the potential and brane tension are arbitrary exponential functions 
of the DBI field. Finally, we summarise in section \ref{sec_summary}. 
\section{Basic Equations} \label{basic-eqn}

We consider a DBI field, $\phi$,  with the following effective 
action \cite{Silverstein:2003hf}:
\beqn
S=-\int d^4 x \sqrt{-g} 
\bigl(
\frac{1}{f(\phi)}
(\sqrt{1 -2 f(\phi) X}-1)
-V(\phi)\bigr),\nonumber\\
\eeqn
where
\beqn
X=-\frac12 g^{\mu \nu} \partial_\mu \phi \partial_\nu \phi.
\eeqn
$V(\phi)$ is a potential
that arises from quantum interactions beween a  D3-brane associated with $\phi$ 
and other D-branes. 
Although a quadratic potential was considered in 
\cite{Silverstein:2003hf}, discussions on the 
exact form of the potential are still ongoing.
$f(\phi)$ is the inverse of the D3-brane tension and contains
geometrical information about the throat in the 
compact internal space. Current proposals for the form of $f(\phi)$ include 
$f(\phi) = \alpha \phi^{-4}$ ($\alpha$ constant),  for the case of an AdS throat.
Another case considered in the literature is a constant function $f(\phi)=\alpha$ \cite{Pajer:2008uy}.
In this paper, we will keep $f(\phi)$ and $V(\phi)$ as
general non-negative functions, i.e.
$V(\phi) \geq 0$ and $f(\phi) \geq 0$.

In a spatially flat Friedmann-Robertson-Walker (FRW)
metric with scale factor $a(t)$, $X=\dot{\phi}^2/2$ for a homogeneous field $\phi$, and it can be shown that
the energy density and pressure of the DBI field
are given by
\begin{eqnarray}
&&
\rho_\phi = \frac{\gamma^2}{\gamma+1}\dot{\phi}^2 + V(\phi),
\nonumber\\
&&p_{\phi}=\frac{\gamma}{\gamma+1}\dot{\phi}^2 - V(\phi),
\end{eqnarray}
where a dot denotes a derivative with respect to the cosmic time, $t$,
and
\begin{eqnarray}
\label{Lorentz}
\gamma \equiv \frac{1}{\sqrt{1-f(\phi) \dot{\phi}^2}},
\end{eqnarray}
which characterises the motion of the brane
and serves as the Lorentz factor.
As in usual special relativity $\gamma \geq 1$
because $f(\phi)$ is non-negative.

If we also take into account a perfect fluid 
with equation of state
$p_m= w_m \rho_m $ $(-1 < w_m < 1)$,
the basic cosmological background equations are given by
\begin{eqnarray}
&&H^2 = \frac13 \biggl[\frac{\gamma^2}{\gamma+1}
\dot{\phi}^2
+ V(\phi) + \rho_m\biggr],
\label{Freidmann}\\
&&
\ddot{\phi} + \frac{3H}{\gamma^2}\dot{\phi}
+\frac{V_{,\phi}}{\gamma^3} + \frac{f_{,\phi}}{2f}
\frac{(\gamma+2)(\gamma-1)}{(\gamma+1)\gamma} \dot{\phi}^2=0,
\label{scalar_eom}\nonumber\\ \\
&&\dot{\rho}_m + 3 H (1+w_m) \rho_m=0,
\label{fluid_eom}
\end{eqnarray}
where the subscript $\phi$ means differentiation with respect
to the field, $H \equiv \frac{\dot{a}}{a}$ is the Hubble parameter
and we have set $8 \pi G =1$ for simplicity.
Eqs.~(\ref{Freidmann}) - (\ref{fluid_eom})
close the system that determines the dynamics.

In order to obtain the late-time attractor behaviour of the system  
we introduce the following set of dynamical variables:
\begin{eqnarray}
x \equiv \frac{\gamma \dot{\phi}}{\sqrt{3 (\gamma+1)}H},
\;\;\;\;
y \equiv \frac{\sqrt{V}}{\sqrt{3}H},
\;\;\;\;
\tilde{\gamma} \equiv \frac{1}{\gamma}\label{defn_xy}\,.
\end{eqnarray}

By defining $x$ and $y$ in this way it is clear that in the limit of 
$\gamma \to 1$,  we recover the dynamical variables 
for a canonical scalar field
originally proposed in Ref.~\cite{Copeland:1997et}.
By doing this, we can intuitively think of
$x$ corresponding to the contribution of the kinetic energy, 
and $y$ to the potential energy of the field. 
For another dynamical degree of freedom which arises due to the 
introduction of the function $f(\phi)$ in this set-up, we adopt $\tilde{\gamma}$
as our third dynamical variable.
This particular choice is intended to make the phase space
being compact.

In terms of these variables,
the Friedmann constraint (\ref{Freidmann})
can be expressed as
\begin{eqnarray}
\label{constraint}
x^2 + y^2 + \Omega_m =1,
\end{eqnarray}
where $\Omega_m \equiv \rho_m /(3 H^2)$.
The energy fraction and equation of state of the DBI field are given by
\begin{eqnarray}
\label{omega_phi}
&&\Omega_\phi = x^2 + y^2,\\
\label{w_phi}
&&w_\phi = \frac{\tilde{\gamma} x^2 - y^2}{x^2 + y^2}.
\end{eqnarray}

Similarly, for a general potential 
$V(\phi)$ and brane tention $f(\phi)$, by introducing
$\lambda$ and $\mu$ which are defined by 
\begin{eqnarray}
&& \lambda \equiv -\frac{V_{,\phi}}{V},
\;\;\;\;\;\;\;\;\;\;\;\;
\mu \equiv -\frac{f_{,\phi}}{f},
\label{defn_lambdamu}
\end{eqnarray}
Eqs.~(\ref{scalar_eom}) 
and (\ref{fluid_eom}) can be written as
\begin{eqnarray}
x_{,N}&=&
\frac{1}{2}\sqrt{3  \tilde{\gamma} (1+ \tilde{\gamma})}
\lambda y^2 
\nonumber\\
&&+ \frac{3}{2} x 
[(1+ \tilde{\gamma}) (x^2-1) +(1+w_m)(1-x^2-y^2)] 
,\label{x_evol_eq_exp}\nonumber\\
\\
y_{,N}&=&-\frac{1}{2}\sqrt{3 \tilde{\gamma}
(1+\tilde{\gamma})} \lambda x y \nonumber\\
&& + \frac{3}{2} y [(1+\tilde{\gamma}) x^2 + 
(1+w_m)(1-x^2-y^2)],
\label{y_evol_eq_exp}\\
\tilde{\gamma}_{,N}&=&
\frac{\tilde{\gamma}(1-\tilde{\gamma}^2)}
{\sqrt{1+\tilde{\gamma} }}
\left[3 \sqrt{1+\tilde{\gamma}} +
\sqrt{3 \tilde{\gamma}}
\frac{1}{x}(\mu  x^2 - \lambda y^2)\right]
,\label{gamma_evol_eq_exp}\nonumber\\
\lambda_{,N} &=& -\sqrt{3 \tilde{\gamma} (1+\tilde{\gamma})}
\lambda^2 (\Gamma -1) x\,,
\label{lambda_evol_eq}\\
\mu_{,N} &=& -\sqrt{3 \tilde{\gamma} (1+\tilde{\gamma})}
\mu^2 (\Xi -1) x \,,
\label{mu_evol_eq}
\end{eqnarray}
where $N \equiv \ln a$, $x_{,N} \equiv \frac{dx}{dN}$ etc...
and we have introduced $\Gamma$ and $\Xi$ through
\begin{eqnarray}
\Gamma  \equiv \frac{V V_{,\phi \phi}}{V_{,\phi}^2}\,,
\;\;\;\;
\Xi  \equiv \frac{f f_{,\phi \phi}}{f_{,\phi}^2}\,.
\label{defn_Gamma_Xi}
\end{eqnarray}

Although Eqs.~(\ref{x_evol_eq_exp}) - (\ref{mu_evol_eq})
hold for a general potential and brane tension,
if both are given as exponential functions of $\phi$,
$V \propto \exp[-\lambda \phi]$, $f \propto \exp [-\mu \phi]$,
then from Eqs.~(\ref{lambda_evol_eq}) and (\ref{mu_evol_eq}), 
$\lambda$ and $\mu$ become constant and 
Eqs.(\ref{x_evol_eq_exp}) - (\ref{gamma_evol_eq_exp})
form the closed autonomous system. We will see the late-time attractor
structure of this case in Sec.~\ref{section_exp}.

There is an important new feature which emerges in the DBI case and is not 
present for the case of the canonical scalar field. In the latter case 
only an exponential potential can truly lead to an autonomous system, but 
as we shortly show, this fact does not apply
to the DBI field.  By combining the degree of freedoms
related with $V(\phi)$ and $f(\phi)$,
we will be able to obtain an autonomous system for a wider class 
of potentials and brane tensions (see also \cite{Guo:2008sz}).

\section{Models with a general
power-law potential and brane tension
\label{section_powerlaw}}

Here, we consider  models where the scalar potential
and the brane tension are arbitrary non-negative power-law functions
of the DBI field

\begin{eqnarray}\label{Vandf}
V(\phi) = \sigma |\phi|^p,\;\;\;\;\;\;
f(\phi) = \nu |\phi|^r,
\end{eqnarray}
with constants $\sigma >0$ and $\nu >0$.

Guo and Ohta \cite{Guo:2008sz} analysed such a model for the case $p=2$ and $r=-4$ but as we will now show it can be addressed as an autonomous system for more general combinations (the exception being $r+p=0$ which requires a separate treatment as we will see).

\subsection{Autonomous System}

In order to represent Eqs.~(\ref{Freidmann})-(\ref{fluid_eom}) by an autonomous set of equations,
we first introduce the variables $\tilde{\lambda}$
and $\tilde{\mu}$ defined by
\begin{eqnarray}
&&\tilde{\lambda} \equiv -\frac{V_{,\phi}}
{f^{q} 
V^{q+1}}, \hspace{1cm}
\tilde{\mu} \equiv -\frac{f_{,\phi}}
{f^{q+1} V^{q}},
\label{def_till_tilm}
\end{eqnarray}
where $q \equiv -1/(p+r)$. Initially we assume $p+r \neq 0$, and will address that special case later.
Given the definitions in Eqn.~(\ref{def_till_tilm}), we can easily verify that $\tilde{\lambda}$ and $\tilde{\mu}$ are constants given by 
\begin{eqnarray}
\tilde{\lambda} = 
\frac{-\varepsilon p}{\sigma^q \nu^q},
\hspace{1cm}
\tilde{\mu} = 
\frac{-\varepsilon r}{\sigma^q \nu^q}
=\frac{r}{p} \tilde{\lambda},
\label{till_tilm_stringsetup}
\end{eqnarray}
where $\varepsilon$ is $1$ for $\phi >0$ and is $-1$
when $\phi <0$. The requirement that $\tilde{\lambda} >0$ is equivalent
to demanding  $\phi < 0$ for $p>0$, and 
$\phi > 0$ for $p<0$. We, therefore, restrict our analysis to $\tilde{\lambda}>0$ without loss of generality.
Then, for physically interesting cases,
we can also restrict our solutions to those where $\dot{\phi} \geq 0$
($x \geq 0$).

From the form of $V(\phi)$ and $f(\phi)$ in Eq.~(\ref{Vandf}) it follows that $\tilde{\lambda}$ and $\tilde{\mu}$ are related with $\lambda$ and $\mu$ through 
\begin{eqnarray}
&&\lambda = \left(\frac{(1-\tilde{\gamma}) y^2}{\tilde{\gamma}
(1+\tilde{\gamma}) x^2}\right)^q \tilde{\lambda}\,,\;\;\;\;\;\;
\mu =
\left(\frac{(1-\tilde{\gamma}) y^2}{\tilde{\gamma}
(1+\tilde{\gamma}) x^2}\right)^q 
\tilde{\mu}\,.\nonumber\\
\label{rel_till_tilm}
\end{eqnarray}
Therefore since $\lambda_{,N}$ and $\mu_{,N}$ can then be  expressed
in terms of $x_{,N}$, $y_{,N}$, $\tilde{\gamma}_{,N}$,
defined in Eqs.~(\ref{x_evol_eq_exp}) - (\ref{mu_evol_eq}),
we only need to solve those three equations. 

In terms of the dynamical variables we have defined above,
the basic equations can be expressed as 
\begin{eqnarray}
x_{,N} &=& 
\frac{\sqrt{3(1+\tilde{\gamma})}}{2}\tilde{\lambda} \tilde{\gamma}^{\frac12-q} (1-\tilde{\gamma})^q  \frac{y^{2q+2}}{x^{2q}} \nonumber\\
&&+\frac{3}{2}x \biggl[(1+\tilde{\gamma}) (x^2-1) + (1+w_m)(1-x^2-y^2)\biggr],\nonumber\\
\label{x_evol_eq_mod}\\
y_{,N}&=&-\frac{\sqrt{3(1+\tilde{\gamma})}}{2}\tilde{\lambda}
\tilde{\gamma}^{\frac12-q}
(1-\tilde{\gamma})^q 
\frac{y^{2q+1}}{x^{2q-1}}
\nonumber\\
&&+\frac{3}{2}y 
\bigl[(1+\tilde{\gamma})x^2 + (1+w_m)(1-x^2-y^2)\bigr],
\label{y_evol_eq_mod}\\
\tilde{\gamma}_{,N}
&=& 
\tilde{\gamma} (1-\tilde{\gamma}^2)\times
\nonumber\\
&&
\left[3+\sqrt{\frac{3}{1+\tilde{\gamma}}} \tilde{\lambda}
\tilde{\gamma}^{\frac{1}{2} -q} (1-\tilde{\gamma})^q
\frac{y^{2q}}{x^{2q+1}}
\left(\frac{r}{p} x^2 -y^2\right)\right].\nonumber\\
\label{gamma_evol_eq_mod}
\end{eqnarray}

Equipped with the basic equations forming an autonomous
system, we will peform the stability analysis 
to obtain the late-time attractor structure 
in the following.

\subsection{Standard Fixed-point Solutions}
\label{subsec_pow_fixed}

Here we obtain fixed-point solutions of the dynamical system given by 
Eqs.~(\ref{x_evol_eq_mod})-(\ref{gamma_evol_eq_mod}). The fact that
there can be terms involving $x,\,y$ and $\tilde{\gamma}$ in the
denominators of these equations means that we have to treat the cases
where these terms vanish with considerable care. In this subsection we
make sure we are working in regimes where any possible ambiguities
involving possible divisions by zero or ratios of `0/0' are avoided. 
In order to distinguish between them, we call the fixed-point solutions
considered in this subsection to be `standard'
fixed-point solutions.
We will address the other cases in subsection \ref{fixedpoints0/0}.

From Eq.~(\ref{gamma_evol_eq_mod}), the requirement that
$\tilde{\gamma}$ is a constant at the standard
fixed points implies three possible scenarios:
\begin{eqnarray}
&&(a)\;\;\tilde{\gamma}=0\,, \nonumber\\
&&(b)\;\;\tilde{\gamma}=1\,, \nonumber\\
&&(c)\;\;  
3=\sqrt{\frac{3}{1+\tilde{\gamma}}} \tilde{\lambda}
\tilde{\gamma}^{\frac{1}{2} -q} (1-\tilde{\gamma})^q
\frac{y^{2q}}{x^{2q+1}}
\left(-\frac{r}{p} x^2 +y^2\right)\,.\nonumber\\
\label{gamma_const_condition}
\end{eqnarray} 

We will investigate each scenario in turn in what follows.

\subsubsection{Case a :  $\tilde{\gamma}=0$}
This result is only valid as long as $q \leq 1/2$.
In this case, since the Lorentz factor (\ref{Lorentz}) tends to infinity,
we shall refer to this class of fixed-points as the ``ultra-relativistic'' solutions. 
Substituting $\tilde{\gamma}=0$ into Eqs.~(\ref{x_evol_eq_mod})-(\ref{gamma_evol_eq_mod}) and eliminating the terms that obviously vanish we obtain
 
\begin{eqnarray}
x_{,N}&=&\frac{\sqrt{3}}{2} \tilde{\lambda} 
\tilde{\gamma}^{\frac{1}{2}-q}
\frac{y^{2q+2}}{x^{2q}}\nonumber\\
&&
+\frac{3}{2} x [(x^2-1)+(1+w_m)(1-x^2-y^2)]\,,
\label{x_evol_eq_mod_a}\\
y_{,N}&=&-\frac{\sqrt{3}}{2} \tilde{\lambda}
\tilde{\gamma}^{\frac{1}{2}-q}
 \frac{y^{2q+1}}{x^{2q-1}}\nonumber\\
&&
+\frac{3}{2} y [x^2+(1+w_m)(1-x^2-y^2)]\,,
\label{y_evol_eq_mod_a}\\
\tilde{\gamma}_{,N} &=& \sqrt{3} \tilde{\lambda}
\tilde{\gamma}^{\frac{3}{2}-q}
\frac{y^{2q}}{x^{2q+1}} 
\left(\frac{r}{p} x^2 -y^2\right)\,.
\label{gamma_evol_eq_mod_a}
\end{eqnarray} 

Due to the freedom in  choosing the sign and the magnitude of $q$, it is 
clear that there are situations where some of the terms in these equations are
ill-defined. We, therefore, classify our analysis in terms of the value of $q$.

For $0 \leq q \leq 1/2$, we find the following 
standard fixed points:

\begin{eqnarray}
(a1)&&(x,y,\tilde{\gamma}) = (1,0,0) \\
(a2)&&(x,y,\tilde{\gamma}) = (x_0,0,0)  \\
&& \,\,\,\,{\rm with}\,\, 0<x_0<1,\;\;w_m = 0\,.
\end{eqnarray}

We refer to the point $(a1)$ as the ultra-relativistic 
kinetic dominated solution, and to the family of solutions (one for each $x_0$)  
$(a2)$ as the ultra-relativistic kinetic-fluid scaling solution. 
For $q\leq -1/2$, we find the following standard fixed-point solution

\begin{eqnarray}
(a3)&&(x,y,\tilde{\gamma}) = (0,1,0)\,,
\end{eqnarray}
which represents the ultra-relativistic potential dominated solutions. 

There are no $\tilde{\gamma} =0$ solutions for the case, $-1/2 < q<0$, but for the 
special case of $q=1/2$, there exist two other 
standard fixed points in the system
of equations (\ref{x_evol_eq_mod_a})-(\ref{y_evol_eq_mod_a}):

\begin{eqnarray}
(a4)&&\;\;\;\;(x,y,\tilde{\gamma})\nonumber\\
&&=\biggl( \sqrt{\frac{\tilde{\lambda} (\sqrt{\tilde{\lambda}^2+12}-\tilde{\lambda})} {6}},\frac{-\tilde{\lambda} +\sqrt{12+\tilde{\lambda}^2}} {2\sqrt{3}},0\biggr)\nonumber\\
\label{fixed_point_kinetic_pot_stringsetup}
(a5)&&\;\;\;\;(x,y,\tilde{\gamma})\nonumber\\
&&=\biggl( \sqrt{-\frac{3(1+w_m)^3}{\tilde{\lambda}^2 w_m}}, \frac{\sqrt{3}(1+w_m)}{\tilde{\lambda}},0 \biggr)\nonumber\\
&&{\rm with}\;\;w_m < 0, \tilde{\lambda} 
\geq \frac{\sqrt{3}(1+w_m)}{\sqrt{-w_m}}\,,
\label{fixed_point_kinetic_pot_stringsetup}
\end{eqnarray}
which we call the ultra-relativistic kinetic-potential scaling solutions,
and the ultra-relativistic kinetic-potential-fluid scaling solutions,
respectively.

It is worth commenting on the solutions $(a1)-(a5)$.  $(a1)$ is well known and corresponds to the case where the kinetic energy of the DBI field dominates over the potential energy, leading to an effective equation of state for the DBI field in Eq.~(\ref{w_phi}) which mimics that of dust ($w_\phi =0$).  The existence of the solution $(a2)$ where this ultrarelativistic kinetic dominated DBI field scales with matter was previously obtained  by Ahn. et. al \cite{Ahn:2009hu,Ahn:2009xd} who also obtained  the solutions (a3) and (a4). In fact $(a3)$ is the solution actively investigated
in the context of DBI inflation, for example as seen in 
Refs.~\cite{Silverstein:2003hf,Alishahiha:2004eh,Chen:2004gc,Chen:2005ad,Shandera:2006ax}. The solution $(a5)$ is of a new type, where the ratio of the kinetic and potential terms
remain constant in the ultrarelativistic limit. The particular case of this solution for  $p=2$ and $r=-4$ was 
first discovered in \cite{Guo:2008sz}, although here we have shown that 
this type of solution exists as long as $p+r=-2$.

\subsubsection{Case b :  $\tilde{\gamma}=1$}

In this case the Lorentz factor $\gamma =1$ and the DBI field mimics the behaviour of a canonical scalar field.
We shall refer to this class of fixed-points as the ``standard'' solutions. 
Substituting $\tilde{\gamma}=1$ into Eqs.~(\ref{x_evol_eq_mod}) - (\ref{gamma_evol_eq_mod}),
we obtain 

\begin{eqnarray}
x_{,N}&=&\frac{\sqrt{6}}{2}\tilde{\lambda} (1-\tilde{\gamma})^q  \frac{y^{2q+2}}{x^{2q}}\nonumber\\
&&
+\frac{3}{2} x [2(x^2-1)+(1+w_m)(1-x^2-y^2)]\,, \label{x_evol_eq_mod_b}\nonumber\\
\\
y_{,N}&=&-\frac{\sqrt{6}}{2} \tilde{\lambda} (1-\tilde{\gamma})^q \frac{y^{2q+1}}{x^{2q-1}}\nonumber\\
&&
+\frac{3}{2} y [2 x^2+(1+w_m)(1-x^2-y^2)]\,, \label{y_evol_eq_mod_b}\\
\tilde{\gamma}_{,N} &=& \sqrt{6} \tilde{\lambda} (1-\tilde{\gamma})^{q+1} \frac{y^{2q}}{x^{2q+1}}  \left(\frac{r}{p} x^2 -y^2\right)\,.
\label{gamma_evol_eq_mod_b}
\end{eqnarray} 

Following our previous approach we classify 
the standard fixed points of this system 
based on the range of values the parameter $q$ can take. We find that for 
$q \geq 0$,  there exists a standard fixed point

\begin{eqnarray}
(b1)\;\;\;\;(x,y,\tilde{\gamma}) = \left(1,0, 1\right)\,,
\end{eqnarray}
which is the standard kinetic energy dominated solution.

For the special case of $q=0$ ($p+r \to \infty$), we find  two 
extra interesting standard  fixed-points:
\begin{eqnarray}
\label{b2}
(b2)\;\;\;\;&&(x,y,\tilde{\gamma}) = \left(\frac{\tilde{\lambda}}{\sqrt{6}}, \frac{\sqrt{6-\tilde{\lambda}^2}}{\sqrt{6}},1 \right),\nonumber\\
&&{\rm with}\;\;\tilde{\lambda} < \sqrt{6}, 
\end{eqnarray}
and 
\begin{eqnarray}
\label{b3}
(b3)\;&&\left(\frac{\sqrt{3} (1+w_m)}{\sqrt{2} \tilde{\lambda}}, \frac{\sqrt{3(1-w_m^2)}}{\sqrt{2} \tilde{\lambda}}, 1 \right),\nonumber\\
&&{\rm with}\;\;\tilde{\lambda} \geq \sqrt{3 (1+w_m)}\;{\rm and}\;-1\leq w_m \leq 1,\nonumber\\\ 
\end{eqnarray}
which we call the standard kinetic-potential scaling solutions,
and the standard kinetic-potential-fluid scaling solutions, 
respectively.

In reviewing these solutions,  recall that the  DBI field behaves just as the usual canonical scalar field
when $\tilde{\gamma} =1$, hence the fixed point solutions found in Case $(b)$  are already well known. For example the properties of $(b2)$ and $(b3)$ 
are identical with that of a standard power-law inflationary solution
and scaling solution respectively, obtained with
an exponential potential in the presence of a canonical scalar field
\cite{Lucchin:1984yf,Ratra_Peebles,trac,Ferreira:1997hj,Copeland:1997et}. In fact $(b2)$ was obtained previously in the context of DBI Inflation in \cite{Ahn:2009hu,Ahn:2009xd}.

\subsubsection{Case c :  $3=\sqrt{\frac{3}{1+\tilde{\gamma}}} \tilde{\lambda}
\tilde{\gamma}^{\frac{1}{2} -q} (1-\tilde{\gamma})^q
\frac{y^{2q}}{x^{2q+1}}
\left(-\frac{r}{p} x^2 +y^2\right)$}

Here $\tilde{\gamma}$ is a constant which is different from $0$ or $1$.
In this case, since the Lorentz factor $\gamma > 1$ and is constant as defined in (\ref{Lorentz}), 
we shall refer to this class of fixed point solutions as the ``relativistic'' ones. 

As there are no values  of $q$ which permit  standard 
fixed-point solutions  $x=y= 0$ in equations (\ref{x_evol_eq_mod})-(\ref{gamma_evol_eq_mod}),
we therefore begin by exploring the possibilities of either $x$ or $y$
being $0$ in this case. 

For the case of $x \neq 0,\; y=0$,
we find the following standard fixed points in the system
which exists only for $q=0$:

\begin{eqnarray}
(c1)&&\;\;\;\;(x,y,\tilde{\gamma}) = \left(1,0, \frac{3}{\tilde{\mu}^2 -3} \right), \nonumber\\
&&{\rm with}\;\; \tilde{\mu} < -\sqrt{6}
\end{eqnarray}
and 
\begin{eqnarray}
(c2)&&\;\;\;\;(x,y,\tilde{\gamma}) = \left(- \frac{\sqrt{3(1+w_m)}} {\sqrt{w_m} \tilde{\mu}},0, w_m \right), \nonumber\\
&&{\rm with}\;\;0 < w_m < 1, \;\; \tilde{\mu} < -\frac{\sqrt{3 (1+w_m)}}{\sqrt{w_m}}, 
\end{eqnarray}
where we refer to $(c1)$ as the relativistic kinetic dominated solution, and to $(c2)$ as the relativistic kinetic-fluid scaling solution.

For the case $x=0,\; y \neq 0$, a standard fixed-point
solution exists but only for $q=-1/2$ 
\begin{eqnarray}
(c3)&&\;\;\;\;(x,y,\tilde{\gamma}) = \left(0, 1, \frac{\sqrt{3}}{\sqrt{\tilde{\lambda}^2 + 3}} \right)\,,
\end{eqnarray}
which we call the relativistic potential dominated late-time solution.

For the remaining cases with $x\neq 0,\; y \neq 0$
it follows that combining
Eqs.~(\ref{x_evol_eq_mod})-(\ref{y_evol_eq_mod}) for general $q$, yields the condition 

\begin{eqnarray}
\label{xneq0ANDyneq0}
3 = \sqrt{\frac{3}{1+\tilde{\gamma}}} \tilde{\lambda} \tilde{\gamma}^{\frac12-q} (1-\tilde{\gamma})^q \frac{y^{2q}}{x^{2q+1}} (x^2+y^2)\,.
\end{eqnarray}
Comparing this with  Eq.~(\ref{gamma_const_condition}), we see that in
Case $(c)$ the condition for a fixed point for general non-zero $x$ and
$y$ requires $p=-r$ (or $q \to -\infty$) a limit we have decided not
analyse in this section.

Summarising the results of Case $(c)$ 
we note that the fixed-point solutions $(c1)$ and $(c2)$
are completely new, while $(c3)$ can also be 
derived as a special case of line 5 of Table~I 
in \cite{Ahn:2009xd}. 
Of particular note for cosmology is  
$(c3)$ which is the concrete example of 
an inflationary solution 
with constant $\tilde{\gamma}$ which differs from  $1$ and $0$. 
Phenomenologically this is a very interesting solution
when considering cosmological perturbations in DBI inflation.

In TABLE~\ref{fixed-points-Summary_power}. we have provided a breakdown
of the $11$ standard fixed-point solutions obtained in these class of
models corresponding to cases $(a)$ - $(c)$.

\subsection{Stability Analysis for standard fixed-points}
\label{subsec_half}

We now turn our attention to carrying out a stability analysis 
for the standard fixed-points obtained in the previous section. Calling
these points in general $x_c,\,y_c$ and 
$\tilde{\gamma}_c$,
we consider small fluctuations about them given by

\begin{eqnarray} 
x = x_c + \delta x\,, \hspace{1cm} y = y_c + \delta y\,,
\hspace{1cm} \tilde{\gamma} = \tilde{\gamma}_c + \delta \tilde{\gamma},\nonumber\\
\end{eqnarray} 
and consider solutions of the form $\delta x \propto e^{w N}$, $\delta y \propto e^{w N}$ and $\delta \tilde{\gamma} \propto e^{w N}$.
As before we consider each case in turn and examine the dynamical behaviour of the system 
close to their fixed-point positions on the phase plane. 

\subsubsection{Case a :  $\tilde{\gamma}=0$}

Expanding Eqs.~(\ref{x_evol_eq_mod})-(\ref{gamma_evol_eq_mod}) 
around the ultra-relativistic kinetic energy dominated solution $(a1)$,  the ultra-relativistic kinetic-fluid scaling solutions $(a2)$ and the ultra-relativistic potential dominated solution $(a3)$ we obtain 
the following eigenvalues 

\begin{eqnarray}
(a1): &&w_1 = 3\,,\;\;
w_2 = \frac32\,,\;\;
w_3 = -3 w_m\, \label{a1-stability} \\
(a2): &&w_1 = 3\,,\;\;
w_2 = \frac32\,,\;\; 
w_3 = 0\,\label{a2-stability} \\
(a3): &&w_1 = -\frac32\,,\;\;
w_2 = -3(1+w_m)\,,\;\; 
w_3 = 3\,. \label{a2-stability} 
\end{eqnarray}
The presence of positive eigenvalues in each case indicates that all three solutions are always unstable. 

For the ultra-relativistic kinetic-potential scaling solution $(a4)$, we obtain the following eigenvalues
\begin{eqnarray}
&& w_{1,2} =  \frac{3 (4 w_m + 8 +\tilde{\lambda}^2 - \tilde{\lambda} \sqrt{12 + \tilde{\lambda}^2})}{8}\nonumber\\
&& \times \biggl[-1 \pm \sqrt{1-\mathcal{A}(\tilde{\lambda}, w_m)}\biggr]\nonumber\\
&&w_3 = \frac{1}{2}\left(-\tilde{\lambda} + \sqrt{12+\tilde{\lambda}^2} \right) (\tilde{\lambda}+\tilde{\mu}) \nonumber\\
&&
\;\;\;\;{\rm where} \;\;\;\;
\mathcal{A}(\tilde{\lambda}, w_m)=8 \left(-12 - \tilde{\lambda}^2 + \tilde{\lambda}\sqrt{12 + \tilde{\lambda}^2}\right) \nonumber\\
&&\times\frac{-6 w_m -6 -\tilde{\lambda}^2 + \tilde{\lambda} \sqrt{12 + \tilde{\lambda}^2}}{9(-\tilde{\lambda}^2 + \tilde{\lambda} 
\sqrt{12 + \tilde{\lambda}^2}-4(2+w_m))^2}\,.\nonumber\\
\label{def_mathcal_a}
\end{eqnarray}

It can be seen that $w_{1,2}$ are non-positive if the condition 
$w_m \geq -1+(\tilde{\lambda}/6)
(\sqrt{\tilde{\lambda}^2 + 12}-\tilde{\lambda})$ is satisfied. Rewriting we see that this condition becomes 
$0<\tilde{\lambda} < \sqrt{3} (1+w_m)/\sqrt{-w_m}$ if $w_m < 0$.
Therefore, considering also $w_3$ we see that these solutions are stable if the additional condition 
$\tilde{\lambda}+\tilde{\mu}<0$ is also satisfied. However from
Eq.~(\ref{till_tilm_stringsetup}) we know that
$\tilde{\lambda}+\tilde{\mu} = \frac{\tilde{\lambda}}{p}(r+p)$ and given
that the condition for the existence of the solution to 
$(a4)$ is $q=1/2$,
we see that this solution is stable only for the case $p>0$ where
$r+p=-2$. 

A similar analysis of the ultra-relativistic kinetic-potential-fluid scaling solution $(a5)$ yields
\begin{eqnarray}
&&w_{1,2} =\frac34 (1-w_m) \left[-1 \pm \sqrt{1+\mathcal{B}(\tilde{\lambda},w_m)}\right]\nonumber\\
&&w_3=3(1+w_m)\frac{\tilde{\lambda}+\tilde{\mu}}{\tilde{\lambda}}\nonumber\\
&&{\rm where}\;\;
\mathcal{B}(\tilde{\lambda},w_m)=\frac{8 (w_m \tilde{\lambda}^2 + 3 (1+w_m)^2)}
{\tilde{\lambda}^2 w_m^2 (1+w_m)}\,.
\label{def_mathcal_b}
\end{eqnarray}

In this case $w_m < 0$ and $\mathcal{B} < 0$ for
$\tilde{\lambda} \geq \sqrt{3} (1+w_m)/\sqrt{-w_m}$,
hence $w_{1,2}$ are negative whenever these solutions exist. 
Of course the bound on $\tilde{\lambda}$ is the complement of that
arising above for the stability of $(a4)$, hence it follows that only
one of the two solutions 
$(a4)$ or $(a5)$ can be stable for a given value of $\tilde{\lambda}$.
The other condition required for stability is once again that 
$w_3 \leq 0$,
which as before corresponds to, $\tilde{\lambda}+\tilde{\mu}<0$ or $p>0$ (recall we 
are assuming $\tilde{\lambda}>0$). It is worth noting that $\tilde{\lambda} = |p|/
\sqrt{\sigma \nu}$ for $q=1/2$.

\subsubsection{Case b :  $\tilde{\gamma}=1$}

For the standard kinetic energy dominated solution $(b1)$ the eigenvalues are
\begin{eqnarray}
w_1 = 3(1-w_m)\,,\;\;
w_2 = 3 \,,\;\;
w_3 = -6 \,,
\end{eqnarray}
for $q \neq 0$ and 
\begin{eqnarray}
w_1 = 3(1-w_m)\,,\;\;
w_2 = 3 -\frac{\sqrt{6}}{2} \tilde{\lambda}\,,\;\;
w_3 = -6-\sqrt{6} \tilde{\mu}, \nonumber\\\,
\end{eqnarray}
for $q = 0$\,,
which clearly indicates that these are unstable solutions. 
For  the standard kinetic-potential scaling solutions $(b2)$, 
(also $q=0$), we find the eigenvalues
\be
w_1 = \frac{\tilde{\lambda}^2-6}{2},\,\,\,
w_2 = \tilde{\lambda}^2-3(1+w_m),\,\,\,
w_3 = -\tilde{\lambda} (\tilde{\lambda}+\tilde{\mu})\,,
\ee
suggesting these solutions are stable when 
$\tilde{\lambda} < \sqrt{3 (1+w_m)}$ and 
$\tilde{\lambda}+\tilde{\mu} > 0$. It is worth noting that $\tilde{\lambda}= |p|$ and 
$\tilde{\lambda} + \tilde{\mu} = -|p|/(pq)$ for
$q=0$. Then, the condition $\tilde{\lambda}+\tilde{\mu} > 0$
corresponds to $p<0$ for $q \to +0$ and $p>0$ for
$q \to -0$.

Similarly, for the standard kinetic-potential-fluid scaling solutions $(b3)$
we obtain 
\begin{eqnarray}
&&w_{1,2} = \frac{3}{4}(1-w_m)\nonumber\\
&&\times \biggl[-1 \pm \sqrt{1-\frac{8 (3 (1+w_m) - \tilde{\lambda}^2) (w_m -1)(1+w_m)}{\tilde{\lambda}^2 (-1+w_m)^2}}\biggr] \nonumber\\
&&w_3 = -3(1+w_m)\frac{\tilde{\lambda}+\tilde{\mu}}{\tilde{\lambda}}\,.
\end{eqnarray}

Since $\tilde{\lambda} \geq \sqrt{3(1+w_m)}$ whenever these solutions exist, $w_1$ and $w_2$ are always negative.
Therefore, these solutions are stable if the condition $\tilde{\lambda}+\tilde{\mu} > 0$  is satisfied. Note also from the condition above for the stability of $(b2)$, if the conditions for both $(b2)$ and $(b3)$ occur, only $(b3)$ will be the stable late time attractor, in other words the standard kinetic-potential-fluid scaling solution will dominate over the standard kinetic-potential scaling solution.

\subsubsection{Case c :  $3=\sqrt{\frac{3}{1+\tilde{\gamma}}} \tilde{\lambda}
\tilde{\gamma}^{\frac{1}{2} -q} (1-\tilde{\gamma})^q
\frac{y^{2q}}{x^{2q+1}}
\left(-\frac{r}{p} x^2 +y^2\right)$}

For the relativistic kinetic energy dominated solution $(c1)$, expanding around this fixed point yields 
the folowing eigenvalues
\begin{eqnarray}
&&w_1 = 3(-w_m +\tilde{\gamma})\nonumber\\
&& w_2 = \frac{3 + 3 \tilde{\gamma} 
- \sqrt{3} \tilde{\lambda} 
\sqrt{\tilde{\gamma} (1+\tilde{\gamma} )}}{2}
\nonumber\\
&&w_3 = \frac{3(-1+\tilde{\gamma})}{2}\,,
\end{eqnarray} 
where $w_3$ is clearly always negative. The parameter space for which
this solution is stable is 
$\tilde{\lambda} \geq \sqrt{3 (1+\tilde{\gamma})/\tilde{\gamma}}$ and $w_m \geq \tilde{\gamma}$.
This is natural since if the potential is steep enough, the kinetic term easily dominates the potential term
and if $w_m \geq \tilde{\gamma} (=w_\phi)$, the energy density
of the fluid decreases faster than that of the DBI field even though it is dominated by the kinetic term.
It is also worth noting that since $\tilde{\gamma}=3/(\tilde{\mu}^2-3)$ for this fixed-point,
the stability conditions are given in terms of 
$ \tilde{\mu}$ as $\tilde{\lambda}^2 > \tilde{\mu}^2$  
and $w_m >3/(\tilde{\mu}^2-3)$. We can go a little further using
Eq.~(\ref{till_tilm_stringsetup}) implying 
$\tilde{\lambda} = -\varepsilon p$. It follows that the condition for
stability is $\frac{r^2}{p^2} < 1$ with $\frac{r}{p} <0$, so $r$ and $p$ have to have the opposite sign whilst satisfying $q=0$. 

For the relativistic kinetic-fluid scaling solutions $(c2)$, a similar analysis produces 
\begin{eqnarray}
&&w_{1,2} = \frac{3}{4} (1-w_m)\nonumber\\
&&\times \biggl[-1 \pm \sqrt{1-\frac{8(3(1+w_m)(\tilde{\mu}^2 w_m -3 (1+w_m))) }{ \tilde{\mu}^2 (-1+w_m)}} \biggr]\nonumber\\
&& w_3 = \frac{3(\tilde{\lambda} + \tilde{\mu})(1+w_m)}{2 \tilde{\mu}} \,.
\end{eqnarray}
Since $\tilde{\mu}^2 > 3 (1+w_m)/w_m$ for these solutions, either $w_1$ or
$w_2$ is always positive, which means these solutions are unstable.

The relativistic potential dominated solution $(c3)$
yields the following eigenvalues
\begin{eqnarray}
w_1 = 0 \,,\;\;
w_2 = -3 \,,\;\;
w_3 = -3(1+w_m) \,.
\end{eqnarray}
As this has a $zero$ eigenvalue, we say this is a
`marginally stable' solution in the sense that
 there is no instability growing exponentially,
although it could be unstable 
to higher orders in the perturbation.
Obviously, the stability of this point is weaker than
that of the fixed-point with three negative eigenvalues.

To help clarify all the possible standard fixed-point 
solutions and their stability the information just provided is summarised in Table~\ref{fixed-points-Summary_power} below.

\begin{widetext}
\begin{center}
\begin{table} [h!]
\begin{tabular} { |c|c|c|c|c|c|c|c| }
\hline 
 & $x$ & $y$ & $\tilde{\gamma}$ & $\Omega_\phi$ & Valid $q$ & Existance & Stability \\
\hline
$(a1)$ & $ 1$ & $0$ & $0$ &$1$ &$0\leq q \leq1/2$ & $\forall w_m, \forall  \tilde{\lambda}, \forall \tilde{\mu}$ & unstable\\      
\hline
$(a2)$ & $x_0$ & $0$ & $0$ & $x_0 ^2$ & $0\leq q \leq 1/2$ & $0<x_0<1$, $w_m=0$, $\forall \tilde{\lambda}, \forall  \tilde{\mu}$ & unstable\\ 
\hline
$(a3)$ & $0$ & $1$ & $0$ & $1$ & $q \leq -1/2$ & $\forall w_m, \forall \tilde{\lambda}, \forall \tilde{\mu}$ & unstable\\      
\hline
$(a4)$ & $\sqrt{\frac{\tilde{\lambda} (\sqrt{\tilde{\lambda}^2+12}-\tilde{\lambda})} {6}}$ & $\frac{-\tilde{\lambda} +\sqrt{12+\tilde{\lambda}^2}}
{2\sqrt{3}}$ & $0$ & $1$ & $q=1/2$ & 
$\forall w_m, \forall  \tilde{\lambda}, \forall  \tilde{\mu}$ & $w_m >
 -1+x_c^2,\; p>0$ \\     
\hline
$(a5)$ & $\sqrt{-\frac{3(1+w_m)^3}{\tilde{\lambda}^2 w_m}}$ & $ \frac{\sqrt{3}(1+w_m)}{\tilde{\lambda}}$ & 
$0$ & $-\frac{3(1+w_m)^2}{\tilde{\lambda}^2 w_m}$ & $q=1/2$ & 
$w_m<0, \tilde{\lambda} \geq \frac{\sqrt{3} (1+w_m)}{\sqrt{-w_m}},
 \forall  \tilde{\mu}$ & $\frac{p}{\sqrt{\sigma \nu}} \geq 
\frac{\sqrt{3}(1+w_m)}{\sqrt{-w_m}}$, \\
& & & & & & & $w_m <0$, $p>0$ \\ 
\hline
$(b1)$ & $ 1$ & $0$ & $1$ &$1$ &$q \geq 0$ & $\forall w_m, \forall  \tilde{\lambda}, \forall \tilde{\mu}$ & unstable \\      
\hline
$(b2)$ & $ \frac{\tilde{\lambda}}{\sqrt{6}}$ & $\frac{\sqrt{6-\tilde{\lambda}^2}}{\sqrt{6}}$ & $1$ &$1$ &  
$q = 0$ & $\forall w_m, \tilde{\lambda} < \sqrt{6}, \forall \tilde{\mu}$ & $ |p| < \sqrt{3(1+w_m)}$,\\
& & & & & & & $1+\frac{r}{p} >0 $\\      
\hline
$(b3)$ & $ \frac{\sqrt{3} (1+w_m)}{\sqrt{2} \tilde{\lambda}}$ & $\frac{\sqrt{3(1-w_m^2)}}{\sqrt{2} \tilde{\lambda}}$ 
& $1$ &$\frac{3(1+w_m)}{\tilde{\lambda}^2}$ & $q = 0$ & $\forall w_m,
 \tilde{\lambda} \geq \sqrt{3 (1+w_m)}, \forall \tilde{\mu}$ & $ 
|p|  \geq \sqrt{3(1+w_m)}$, \\ 
& & & & & & & $1+\frac{r}{p} >0 $ \\      
\hline
$(c1)$ & $1$ & $0$ & $\frac{3}{\tilde{\mu}^2-3}$ & $1$ & $q=0$ & $\tilde{\mu}<-\sqrt{6}, \forall w_m, \forall \tilde{\lambda}$ & 
$w_m > \frac{3}{r^2-3} (>0)$,\\
& & & & & & & $0>\frac{r}{p}>-1$  \\      
\hline
$(c2)$ & $-\frac{\sqrt{3(1+w_m)}}{\sqrt{w_m}\tilde{\mu}}$ & $0$ & $w_m$ & $\frac{3(1+w_m)}{w_m \tilde{\mu}^2}$ &
$q=0$ & $0 < w_m <1, \tilde{\mu} < -\frac{\sqrt{3 (1+w_m)}}{\sqrt{w_m}}, \forall \tilde{\lambda}$ & unstable \\      
\hline
$(c3)$ & $0$ & $1$ & $\frac{\sqrt{3}}{\sqrt{\tilde{\lambda}^2 + 3}}$ &
 $1$ & $q=-1/2$ & $\forall w_m, \forall \tilde{\lambda}, \forall
 \tilde{\mu}$ & marginally stable\\      
\hline

\end{tabular}
\caption{Summary of the standard fixed-points and their 
stability in the models where the potential 
and brane tension are arbitrary non-negative power-law functions of the DBI field. Notice that we have restricted 
$\tilde{\lambda} > 0$ and $x \geq 0$. We have expressed the conditions for the stability
in terms of $(p,r)$ not $(\tilde{\lambda}, \tilde{\mu})$
as these are the more fundamental quantities.}
\label{fixed-points-Summary_power}
\end{table}
\end{center}
\end{widetext}

\subsection{\label{fixedpoints0/0}
Fixed-points arising from solutions that initially appear singular}

As mentioned earlier, for models with a general
power-law potential and warp factor, 
in addition to the usual standard fixed points,
it is necessary to check if the points where 
the denominator is $0$  in the right hand side of 
Eqs.~(\ref{x_evol_eq_mod})-(\ref{gamma_evol_eq_mod}) can also 
be late-time attractor solutions for the system,
a situation which does not arise in the case with a canonical scalar field 
\cite{Copeland:1997et}.

Even in the case that the point itself leads to a
singularity, it doesn't necessarily mean the system is ill defined. For example it could be that the  solutions approach the point
exponentially slowly (i.e. like $\exp[-N]$), hence it would take  
an infinite time to reach the singularity
and physically there is no ill behaviour in the system. 
In particular as long as the ratio of the singular terms (loosely called `$0/0$') tends to a constant value then the system can be analyzed for the stability of these fixed points. 
As standard techniques can be applied to judge
the stability of such a point, we also call them
fixed-points in what follows.
In the following, depending on the value of $q$
we show there are 6 kinds of fixed-point where 
 `$0/0$' is finite in the phase space. 
As in the previous analysis, we have excluded
the special case with $p=-r$ here.

\subsubsection{Standard potential dominated solutions}

First, we consider the point 
($\alpha$1):$(x,y,\tilde{\gamma}) = (0,1,1)$ 
which in the case of a canonical field would simply corresponds to a standard slow-roll inflationary
solution. However, in this case $x=0$ leads to a singularity in Eqs.~(\ref{x_evol_eq_mod})-(\ref{gamma_evol_eq_mod}) for $q>-1/2$ and $\tilde{\gamma} =1$ does the same for $q<0$, hence we have to tread carefully in analysing the system.

Since the coordinate $y=1$ does not result in any ill-defined behaviour 
in Eqs.~(\ref{x_evol_eq_mod})-(\ref{gamma_evol_eq_mod}),
we can consider the reduced system in which we determine the leading order behaviour of $x$ and 
$\tilde{\gamma}$ around the point $(\alpha 1)$.
Writing $x = 0 + \delta x$, $\tilde{\gamma} = 1 - \delta \tilde{\gamma}$ and keeping only the leading
order terms, we obtain
\begin{eqnarray}
\label{xalpha1}
\delta x_{,N} &=& 
-\frac{\delta x}{\sqrt{2}} 
\left( 3 \sqrt{2} - \tilde{\lambda}
\frac{\sqrt{3} (\delta \tilde{\gamma})^q}{(\delta x)^{2q+1}}   \right) \,, \\
\label{gammaalpha1}
\delta \tilde{\gamma}_{,N} &=&
 - \sqrt{2} \delta \tilde{\gamma}\left( 3 \sqrt{2} -
\tilde{\lambda} \frac{
\sqrt{3} (\delta \tilde{\gamma})^q}{(\delta x)^{2q+1}}\right) \,.
\end{eqnarray}

If we introduce  $\beta \equiv \frac{(\delta x)^{2q+1}}
{(\delta \tilde{\gamma})^q}$, for $q<-1/2$ and 
$q>0$, this appears to be of the form of `$0/0$' and
requires a careful analysis to properly understand the behaviour of this system. In this case
from  Eqs.~(\ref{xalpha1}) - (\ref{gammaalpha1}),
it becomes
\begin{eqnarray}
\frac{(\delta x)^{2q+1}}
{(\delta \tilde{\gamma})^q} = c e^{-3 N} + 
\frac{\sqrt{6}}{6} \tilde{\lambda}\,,
\label{betaalpha1}
\end{eqnarray}
where $c$ is an integration constant. At late times
($N \to \infty$) we see that even though both terms tend to zero,
the ratio $(\delta \tilde{\gamma})^q / (\delta x)^{2q+1}$
approaches the contant given by 
$(\delta \tilde{\gamma})^q / (\delta x)^{2q+1}
= \sqrt{6}/\tilde{\lambda}$. Then,
 it is possible to sensibly 
discuss whether the point $(\alpha 1) = (0,1,1)$ is stable or not
in terms of the remaining two-dimensional system obtained 
by substituting 
$(1- \tilde{\gamma})^q / x ^{2q + 1} = 
\sqrt{6}/\tilde{\lambda}$  back into 
Eqs.~(\ref{x_evol_eq_mod})-(\ref{y_evol_eq_mod}).
It is worth noting that as the coefficient of $N$
in the exponential
function in Eq.~(\ref{betaalpha1}) is $-3$ which is negative,
this solution is stable along the direction of $\beta$.

The eigenvalues $w_1,\,w_2$ and $w_3$ corresponding to evolution of the perturbations in $x,\,y$ and $\tilde{\gamma}$ respectively are obtained from Eqs.~(\ref{xalpha1}),(\ref{y_evol_eq_mod})
and (\ref{gammaalpha1}):

\be
w_1 = 0=w_3\,, \hspace{1cm} w_2 = -3(1+w_m)\,.
\label{eigen_alpha1}
\ee

To be specific the results $w_1 = 0=w_3$ arise because
at leading order, the analytic solution including
$(\delta \tilde{\gamma})^q / (\delta x) ^{2q + 1}= 
\sqrt{6}/\tilde{\lambda}$
leads to a vanishing right hand side for Eqs.~(\ref{xalpha1})
and (\ref{gammaalpha1}). 
The fact that at leading order all eigenvalues 
are non-positive, implies 
that for $q<-1/2$ and $q>0$ there are solutions which tend to
$(\alpha 1)$ regardless of the values of 
$\tilde{\lambda}$, $\tilde{\mu}$ and $w_m$
even though it is not a fixed-point solution
in a usual sense. 

But this $zero$ eigenvalue implies that
the stability of this point is weaker than
the point which has negative eigenvalues for all three
directions. As in the case $(c3)$, we say this solution is `marginally stable'. 
Of course, to obtain the strict stability of this solution  
we would have to go to higher order. 

\subsubsection{Ultra-relativistic potential dominated solutions}

Next, we consider the point ($\alpha$2):$(x,y,\tilde{\gamma}) = 
(0,1,0)$ whose behaviour is the same as $(a3)$ except for the fact that 
$x=0$ is singular for $q>-1/2$
and $\tilde{\gamma} =0$ 
for $q>1/2$ 
in Eqs.~(\ref{x_evol_eq_mod})-(\ref{gamma_evol_eq_mod}).

Following the arguments used for  ($\alpha 1$)
and writing $x=0+\delta x$, 
$\tilde{\gamma} = 0+\delta \tilde{\gamma}$  
we obtain to leading order:

\begin{eqnarray}
\label{xalpha2}
\delta x_{,N} &=&  \frac{\delta x}{2} \left( -3 + 
\sqrt{3}\tilde{\lambda} \frac{(\delta \tilde{\gamma})^{\frac{1}{2}-q}}
{(\delta x)^{2q+1}} \right) \,,\\
\label{gammaalpha2}
\delta \tilde{\gamma}_{,N} &=&  - \delta \tilde{\gamma} 
\left( -3 + \sqrt{3}\tilde{\lambda} 
 \frac{(\delta \tilde{\gamma})^{\frac{1}{2}-q}}
{(\delta x)^{2q+1}} \right)\,.
\end{eqnarray}

For $-1/2 < q < 1/2$, $\beta \equiv \frac{(\delta x)^{2q+1}}
{(\delta \tilde{\gamma})^{1/2-q}}$ once again appears ill defined, however we can solve 
Eqs.~(\ref{xalpha2}) - (\ref{gammaalpha2}) to give
\begin{eqnarray}
\frac{(\delta x)^{2q+1}}
{(\delta \tilde{\gamma})^{\frac{1}{2}-q}} = c e^{-3 N} + 
\frac{\sqrt{3}}{3} \tilde{\lambda}\,,
\label{betaalpha2}
\end{eqnarray}
where $c$ is an integration constant. At late times
($N \to \infty$) we see that even though both terms tend to zero,
the ratio 
$(\delta \tilde{\gamma})^{1/2-q} / (\delta x)^{2q+1}
\to \sqrt{3}/\tilde{\lambda}$.

As in the case of ($\alpha 1$), it is possible to
discuss whether the point $(\alpha 2) = (0,1,0)$
is stable or not in terms of 
the remaining two-dimensional system 
obtained by substituting the particular solution
$\tilde{\gamma}^{1/2-q}/x^{2q+1} = \sqrt{3}/\tilde{\lambda}$  
back into Eqs.~(\ref{x_evol_eq_mod})-(\ref{y_evol_eq_mod}). 
It is worth noting that as the coefficient of 
$N$ in the exponential
function in Eq.~(\ref{betaalpha2}) is $-3$ 
(i.e. negative),
this solution is stable along the direction of $\beta$.

The corresponding eigenvalues for the perturbations in $x,y$ and $\tilde{\gamma}$ follow from Eqs.~(\ref{xalpha2}), (\ref{y_evol_eq_mod}) and (\ref{gammaalpha2}) and are given by 
\be
w_1 = 0=w_3\,, \hspace{1cm} w_2 = -3(1+w_m)\,.
\ee

Note that the eigenvalues for $(\alpha 2)$ are identical to those of $(\alpha 1$) hence to obtain the full stability of the system, as in that case, we would have to go to higher order
to obtain the strict stability of the system

\subsubsection{Ultra-relativistic kinetic dominated solutions}

Next, we consider the point 
($\alpha$3):$(x,y,\tilde{\gamma}) = (1,0,0)$ 
whose property is the same as $(a1)$ except that 
$y=0$ is singular for $q < 0$
and $\tilde{\gamma} =0$ is singular
for $q>1/2$ 
in Eqs.~(\ref{x_evol_eq_mod})-(\ref{gamma_evol_eq_mod}).

In this case, since the coordinate $x=1$ does not result 
in any ill-defined behaviour 
in Eqs.~(\ref{x_evol_eq_mod})-(\ref{gamma_evol_eq_mod}),
we can consider the reduced system in which we determine 
the leading order behaviour of $y$ and 
$\tilde{\gamma}$ around the point $(\alpha 3)$.
Writing $y = 0 + \delta y$, $\tilde{\gamma} = 0+ \delta \tilde{\gamma}$ and keeping only leading
order terms, we obtain
\begin{eqnarray}
\label{y-alpha3}
\delta y_{,N} &=& 
\frac{\sqrt{3}}{2} \delta y 
\left( \sqrt{3} - \tilde{\lambda}
\frac{(\delta \tilde{\gamma})^{\frac{1}{2}-q}}{(\delta y)^{-2q}}   \right) \,, \\
\label{gammaalpha3}
\delta \tilde{\gamma}_{,N} &=&
\sqrt{3} \delta \tilde{\gamma} \left( \sqrt{3}+\frac{r}{p} 
\tilde{\lambda}\frac{(\delta 
\tilde{\gamma})^{\frac{1}{2}-q}}
{(\delta y)^{-2q}} \right) \,, 
\end{eqnarray}

Following the earlier examples, for $q<0$ and $q>1/2$, we introduce $\beta \equiv 
\frac{(\delta y)^{-2q}}{(\delta \tilde{\gamma})^{1/2-q}}$
and proceed to show that it has a well behaved non-trivial behaviour of its own.
Solving Eqs.~(\ref{y-alpha3})-(\ref{gammaalpha3}) we find
\begin{equation}
\label{betaalpha3}
\frac{(\delta y)^{-2q}}{(\delta 
\tilde{\gamma})^{\frac{1}{2}-q}}
 = ce^{-\frac32 N} + {\sqrt3 \over 3}\tilde{\lambda} 
\left(2 q + (2 q -1){r \over p}\right)\,,
\end{equation}
where $c$ is an integration constant. 
At late times ($N\to \infty$) we see  that even  both terms tend to zero, the ratio 
$(\delta \tilde{\gamma})^{1/2-q}/(\delta y)^{-2q}$ 
approaches the constant given by 
$(\delta \tilde{\gamma})^{1/2-q}/(\delta y)^{-2q} = 
\sqrt{3}/[ \tilde{\lambda}(2q +(2q-1)r/p)]$.
Since both $\delta y$ and $\delta \tilde{\gamma}$
are positive, for this solution to be physical,
for a given $q$, $r/p$ should satisfy
\be
2q+(2q-1)\frac{r}{p} >0\,,
\label{existalpha3}
\ee
which is equivalently $p>0$ with $r<-2$ or $p<0$ with 
$r>-2$.

As in the previous cases, it is possible to
discuss whether the point $(\alpha 3) = (1,0,0)$
is stable or not in terms of the remaining
two-dimensional system obtained by substituting 
$\tilde{\gamma}^{1/2-q}/y^{-2q} = 
\sqrt{3}/[ \tilde{\lambda}(2q +(2q-1)r/p)]$
back into 
Eqs.~(\ref{x_evol_eq_mod})-(\ref{y_evol_eq_mod}).
It is worth noting that as the coefficient of 
$N$ in the exponential
function in Eq.~(\ref{betaalpha3}) is $-3/2$ which is negative,
this solution is stable along the direction of $\beta$.

The eigenvalues for the perturbations in $x,\,y$ and $\tilde{\gamma}$ follow from Eqs.~(\ref{x_evol_eq_mod}),(\ref{y-alpha3}) and (\ref{gammaalpha3}) to give 
\begin{eqnarray}
\label{evx-alpha3}
w_1 &=& -3 w_m\\
\label{evy-alpha3}
w_2 &=& \frac{3 (2q-1)(1+\frac{r}{p})}{2[2q + (2q-1) \frac{r}{p}]} = \frac{3}{2}\left[1+\frac{p}{r+2}\right]\,\\
\label{evg-alpha3}
w_3 &=&  \frac{6q (1+\frac{r}{p})}{[2q + (2q-1) \frac{r}{p}]} = {6 \over 2+r},
\end{eqnarray}
where in Eqs.~(\ref{evy-alpha3})-(\ref{evg-alpha3}) we have made use of
the fact $q=-\frac{1}{r+p}$.
Although $w_2$ and $w_3$ diverge for $r=-2$, 
from Eq.~(\ref{existalpha3}) this case is not physical.

From these eigenvalues, we see that for both cases $q<0$ and $q>1/2$, stable solutions require $w_m>0$ and $r<-2$. Note that in Eq.~(\ref{existalpha3}) this implies $p>0$ for stable solutions. For $q<0$, the additional constraints are $r/p >-1$, whereas for the case $q>1/2$ it is $-(2/p+1) < r/p < -1$.  

\subsubsection{Ultra-relativistic kinetic-fluid scaling solutions}

In this case the behaviour is identical to ($\alpha 3$), 
for the special case with $w_m=0$. However, 
there is another important point 
($\alpha$4):$(x,y,\tilde{\gamma}) = (x_0,0,0)$ 
with $x_0$ satisfying $0<x_0<1$ and 
whose property is the same as $(a2)$  in that there are a family of solutions depending on the value of $x_0$ chosen.  Again $y=0$ appears singular for $q < 0$
and $\tilde{\gamma} =0$  for $q>1/2$ 
in Eqs.~(\ref{x_evol_eq_mod})-(\ref{gamma_evol_eq_mod}).

Following the arguments used for ($\alpha 3$)
and writing $x=x_0$, $y = 0 + \delta y$, 
$\tilde{\gamma} = 0+ \delta \tilde{\gamma}$, we obtain to leading order
\begin{eqnarray}
\label{y-alpha4}
\delta y_{,N} &=& 
\frac{\sqrt{3}}{2} \delta y 
\left( \sqrt{3} - \tilde{\lambda}
x_0 ^{1-2q}
\frac{(\delta \tilde{\gamma})^{\frac{1}{2}-q}}{(\delta y)^{-2q}}   \right) \,, \\
\label{gammaalpha4}
\delta \tilde{\gamma}_{,N} &=&
\sqrt{3} \delta \tilde{\gamma} \left( \sqrt{3}+\frac{r}{p} 
\tilde{\lambda} x_0^{1-2q}
\frac{(\delta \tilde{\gamma})^{\frac{1}{2}-q}}
{(\delta y)^{-2q}} \right) \,, 
\end{eqnarray}

For $q<0$ and $q>1/2$, again introducing $\beta \equiv 
\frac{(\delta y)^{-2q}}{(\delta \tilde{\gamma})^{1/2-q}}$, we can solve Eqns.~(\ref{y-alpha4})-(\ref{gammaalpha4}) 
to give
\begin{equation}
\label{betaalpha4}
\frac{(\delta y)^{-2q}}{(\delta \tilde{\gamma})^{1/2-q}}
 = ce^{-\frac32 N} + {\sqrt3 \over 3}\tilde{\lambda}  
x_0 ^{1-2 q}
\left(2 q + (2 q -1){r \over p}\right)\,,
\end{equation}
where $c$ is an integration constant. 
Once again at late times ($N\to \infty$) we see that  even 
though both terms tend to zero, the ratio 
$(\delta \tilde{\gamma})^{1/2-q}/(\delta y)^{-2q}$ 
approaches the constant given by 
$(\delta \tilde{\gamma})^{1/2-q}/(\delta y)^{-2q} = 
\sqrt{3}/[ \tilde{\lambda} x_0 ^{1-2q} (2q +(2q-1)r/p)]$.
Since both $\delta y$ and $\delta \tilde{\gamma}$
are positive, for this solution to be physical,
for a given $q$, $r/p$ should satisfy 
Eq.~(\ref{existalpha3}). 
It is worth noting that the coefficient of $N$
in the exponential
function in Eq.~(\ref{betaalpha4}) is $-3/2$ 
which being negative implies 
this solution is stable along the direction of $\beta$.

The eigenvalues for the perturbations in $x,\,y$ and $\tilde{\gamma}$ turn out to be identical to those for ($\alpha 3$) in Eqs.~(\ref{evy-alpha3}) and (\ref{evg-alpha3}), except for the case of $w_1$. In particular from Eqs.~(\ref{x_evol_eq_mod}),(\ref{y-alpha4}) and (\ref{gammaalpha4}) we obtain 
\begin{eqnarray}
\label{evx-alpha4}
w_1 &=& 0\\
\label{evy-alpha4}
w_2 &=& \frac{3 (2q-1)(1+\frac{r}{p})}{2[2q + (2q-1) \frac{r}{p}]} = \frac{3}{2}\left[1+\frac{p}{r+2}\right]\,\\
\label{evg-alpha4}
w_3 &=&  \frac{6q (1+\frac{r}{p})}{[2q + (2q-1) \frac{r}{p}]} = {6 \over 2+r},
\end{eqnarray}
Again, although $w_2$ and $w_3$ diverge for $r=-2$, 
this case is excluded as $\beta$ is then no longer a finite constant.

$w_1$ shows that although there is a $zero$ eigenvalue
along the $x$ direction, when $w_2 < 0$, there are solutions
approaching the point ($\alpha 4$). 
From the discussions for ($\alpha 3$), we see that for both cases $q<0$ and $q>1/2$, stable solutions require $r<-2,\, p>0$. For $q<0$, the additional constraints are $r/p >-1$, whereas for the case $q>1/2$ it is $-(2/p+1) < r/p < -1$.  

\subsubsection{Standard fluid dominated solutions}

Next, we consider the point 
($\alpha$5):$(x,y,\tilde{\gamma}) = (0,0,1)$
which in the case of a canonical field corresponds to the usual fluid dominated solution. However, here 
$x=0$ is singular for $q > -1/2$, 
$y=0$ and $\tilde{\gamma} =1$ are singular 
for $q<0$ 
in Eqs.~(\ref{x_evol_eq_mod})-(\ref{gamma_evol_eq_mod}). 
Since  all the coordinates $x=0$, $y=0$, $\tilde{\gamma}=1$
can result in ill defined behaviour in 
Eqs.~(\ref{x_evol_eq_mod})-(\ref{gamma_evol_eq_mod}),
the stability analyis for this point is more complicated than
in the previous examples in this subsection.

To investigate the behaviour of perturbations in 
$x$, $y$ and 
$\tilde{\gamma}$ around $(\alpha 5)$, 
we write $x=0+\delta x$, $y=0+\delta y$,
$\tilde{\gamma}=1-\delta \tilde{\gamma}$
and keep only the leading order terms, 
to obtain 

\begin{eqnarray}
\label{xalpha5}
\delta x_{,N} &=& \frac{ \delta x}{2} 
\left[-3 (1-w_m) + \sqrt{6} \tilde{\lambda}
\frac{(\delta x)^{-1-2q} (\delta y)^{2q + 2}}
{(\delta \tilde{\gamma})^{-q}}\right]\,,\nonumber\\
\\
\delta y_{,N} &=& 
\frac{\delta y}{2}  
\left[ 3(1+w_m) -\sqrt{6} \tilde{\lambda} 
\frac{(\delta x)^{-2q +1}}{(\delta y)^{-2q} 
(\delta \tilde{\gamma})^{-q}} \right]\,,\nonumber\\
\label{yalpha5}\\
\delta \tilde{\gamma}_{,N} &=&-2 \delta \tilde{\gamma}
\left[3+\sqrt{\frac{3}{2}} \tilde{\lambda} 
\frac{(\delta x)^{-2q +1}}{(\delta y)^{-2q} 
(\delta \tilde{\gamma})^{-q}}
\left(\frac{r}{p}- \frac{(\delta y)^2}
{(\delta x)^2}\right)\right]\,.\nonumber\\
\label{gammaalpha5}
\end{eqnarray}

We can in fact make considerable progress by solving 
for combinations of the variables as in the previous example.  
Introducing $\theta \equiv \frac{(\delta x)^{-2q+1}}
{(\delta y)^{-2q} (\delta \tilde{\gamma})^{-q}}$ 
and $\beta = \frac{(\delta y)^2}{(\delta x)^2}$, from Eqns.~(\ref{xalpha5})-(\ref{gammaalpha5})  we obtain

\begin{eqnarray}
\label{thetaalpha5}
\frac{\theta_{,N}}{\theta} &=& -\frac32 (1-w_m) + \sqrt{\frac32}
\tilde{\lambda} \theta \left[
-2q\left(1 + \frac{r}{p}\right)  + \beta \right]\,,\nonumber\\
\\
\label{betaalha5}
\frac{\beta_{,N}}{\beta} &=& 6 - \sqrt{6} \tilde{\lambda} \theta (1+\beta)\,,
\end{eqnarray}

These equations pick up two of the three degrees 
of freedom of the system given by Eqs.~(\ref{xalpha5})
-(\ref{gammaalpha5}), and stability in this two-dimensional
system is necessary for the stability in the full
three-dimensional system. We find this two-dimensional
system has three fixed-points characterised by
\begin{eqnarray}
&&\left(\theta, \beta \right)= \left(0,0\right)\,,\;\;{\rm and}
\label{alpha51}\\
&&
\left(\frac{3(1-w_m)}{-2q \sqrt{6}
 \tilde{\lambda} \left(1+\frac{r}{p}\right)},0\right)
\,,\;\;{\rm and}\label{alpha52}\\
&&\left(\frac{3(1+w_m)}
{\sqrt{6} \tilde{\lambda} \left[2q\left(1+\frac{r}{p}\right)
+1\right]}, \frac{4q \left(1+\frac{r}{p}\right)
+(1-w_m)}{1+w_m}\right)\,.\label{alpha53}\nonumber\\
\end{eqnarray}

It turns out that the eigenvalues corresponding to fluctuations about 
Eq.~(\ref{alpha51}) are $-3(1-w_m)/2$ and $6$
whilst those corresponding to Eq.~(\ref{alpha52})
are $3(1-w_m)$ and $3[(1-w_m) + 4q (1+r/p)]/[2q (1+r/p)]$,
indicating that both solutions are unstable.

Turning to the stability of the fixed-point corresponding to
Eq.~(\ref{alpha53}) we first of all note that in order for the solutions to be physical requires a number of conditions be satisfied. 
For  $\theta$ and $\beta$ to be finite constants requires 
$q<0$ since $\delta x$ and $\delta y$ have the same time dependence. 
Further, since $\delta x$, $\delta y$, $\delta \tilde{\gamma}$ are all positive, 
$\theta$ and $\beta$ must be positive, which leads to the conditions  
\begin{eqnarray}
&&1+2q\left(1 +  \frac{r}{p}\right) >0\,,\label{exist1alpha5}\\
&&4q \left(1+\frac{r}{p}\right) + (1-w_m) >0\,,
\label{exist2alpha5}
\end{eqnarray}
 which are equivalently $p>4/(1-w_m)$ or $p<0$.

Now, by considering small perturbations around this point of the form 
$\delta \theta = A e^{w_a N} + B e^{w_b N},  \delta \beta = C e^{w_a N} + D e^{w_b N}$,
we obtain the following eigenvalues,
\begin{eqnarray}
&&w_{a,b} = \frac{3 (2q(3+w_m)(1+\frac{r}{p}) + (1-w_m))}
{4(1+2q(1+\frac{r}{p}))}\nonumber\\
&&\times\left[-1 \pm
	 \sqrt{1- \mathcal{C} 
(\tilde{\lambda},w_m,\frac{r}{p})}\right]
\nonumber\\
&&{\rm where}\;\mathcal{C} (\tilde{\lambda},w_m,\frac{r}{p})
\;\;
{\rm is}
\nonumber\\
&&
\frac{8(1+2q(1+\frac{r}{p}))(1+w_m)(4q(1+\frac{r}{p})+(1-w_m))}
{(2q(3+w_m)(1+\frac{r}{p}) + (1-w_m))^2}
\,.\nonumber\\
\label{eigen_theta_beta_alpha5}
\end{eqnarray}
and (+,\,-) is associated with $(w_a,\,w_b)$ respectively.
Given the conditions (\ref{exist1alpha5}) and 
(\ref{exist2alpha5})
are satisfied for physically relevant situations, it follows that stable eigenvalues ($w_{a,b} \leq 0$ ) are obtained if 
\begin{eqnarray}
2q(3+w_m)(1+\frac{r}{p}) + (1-w_m) \geq 0\,,
\label{stab1alpha5}
\end{eqnarray}
which is equivalent to $p>2(3+w_m)/(1-w_m)$ or $p<0$.

In fact (\ref{stab1alpha5}) is more
restrictive than (\ref{exist1alpha5}) and (\ref{exist2alpha5})
as long as we are in the physically acceptable range $-1< w_m < 1$.

If the condition (\ref{stab1alpha5}) is satisfied,
the terms including $\delta x$, $\delta y$, 
$\delta \tilde{\gamma}$ in Eqs.~(\ref{xalpha5})
-(\ref{gammaalpha5}) are all finite and constant.
Then it is possible to sensibly discuss whether the
point $(\alpha 5) = (0,0,1)$ is stable or not 
in terms of the remaining one-dimensional system 
obtained by substituding  
(\ref{alpha53}) back into say Eq.~(\ref{x_evol_eq_mod}).

Then we find the eigenvalue for the perturbations in $x$  
\begin{equation}
w_1 = \frac{3q (1+w_m)(1+\frac{r}{p})}{1+2q (1+\frac{r}{p})} = -\frac{3(1+w_m)}{p-2}\,.
\label{eigen_x_alpha5}
\end{equation}
Notice that we have already obtained the stability
for $\theta$ and $\beta$ directions, the stability based on 
the eigenvalues for the directions of $x$, 
$y$ and $\tilde{\gamma}$ have the same information.
Actually, the eigenvalues corresponding to
the perturbation in $y$ and $\tilde{\gamma}$ are given by
\begin{eqnarray}
w_2 &=&   \frac{3q (1+w_m)(1+\frac{r}{p})}{1+2q (1+\frac{r}{p})}=-\frac{3(1+w_m)}{p-2}\,,\\
w_3 &=&   -\frac{3 (1+w_m)(1+\frac{r}{p})}{1+2q (1+\frac{r}{p})}\,,
\end{eqnarray}
which are consistent with
\begin{eqnarray}
&&w_1 =w_2\,,\\
&&(-2q+1)w_1 + 2q w_2 + q w_3 =0\,,
\end{eqnarray}
derived from $\theta$ and $\beta$ are constant.
Although $w_1$, $w_2$ and $w_3$ diverge for $p=2$,
from Eq.~(\ref{stab1alpha5}), this case is excluded.

We can summarise the constraints on the allowed parameters if the solution
 ($\alpha 5$) is to be stable. From $q<0$ we have $r+p>0$
and from $w_1<0$ we have $p>2$ and $1+r/p>0$. Then, from 
Eq.~(\ref{stab1alpha5}), we have 
$p>2(3+w_m)/(1-w_m)>2$ and $w_m <(p-6)/(p+2)$.

\subsubsection{Ultrarelativistic fluid dominated solutions}

Next, we consider the point 
($\alpha$6):$(x,y,\tilde{\gamma}) = (0,0,0)$
which corresponds to a new type of  
fluid dominated solution. We don't consider is as one of the standard fixed points  because of the apparently singular behaviour of the point in various limits. For example in Eqs.~(\ref{x_evol_eq_mod})-(\ref{gamma_evol_eq_mod}), $x=0$ is singular for $q > -1/2$, $y=0$ is singular for $q < 0$ and $\tilde{\gamma} =0$ for $q>1/2$. It complicates the analyis for this point but the procedure itself is similar to the case with ($\alpha$5). 

Writing $x=0+\delta x$, $y=0+\delta y$,
$\tilde{\gamma}=0+\delta \tilde{\gamma}$
and keeping only the leading terms, 
we obtain 

\begin{eqnarray}
\label{xalpha6}
\delta x_{,N} &=& \frac{\delta x}{2} 
\left[3 w_m + \sqrt{3} \tilde{\lambda}
\frac{(\delta y)^{2q + 2}}{(\delta x)^{2q+1} 
(\delta \tilde{\gamma})^{q-\frac12}}
\right]\,,\\
\label{yalpha6}
\delta y_{,N} &=& 
\frac{\delta y}{2}  
\left[ 3(1+w_m) -\sqrt{3} \tilde{\lambda} 
\frac{(\delta y)^{2q}}{(\delta x)^{2q -1}
(\delta \tilde{\gamma})^{q-\frac12}}
\right]\,,\\
\label{gammaalpha6}
\delta \tilde{\gamma}_{,N} &=&\delta \tilde{\gamma}
\left[3+\sqrt{3} \tilde{\lambda} 
\frac{(\delta y)^{2q}}{(\delta x)^{2q -1}
(\delta \tilde{\gamma})^{q-\frac12}}
\left(\frac{r}{p}- \frac{(\delta y)^2}
{(\delta x)^2}\right)\right]\,.\nonumber\\
\end{eqnarray}

Following the previous case ($\alpha 5$),  
we introduce  
$\theta \equiv \frac{(\delta y)^{2q}}{(\delta x)^{2q -1}
(\delta \tilde{\gamma})^{q-\frac12}}$ 
and $\beta = \frac{(\delta y)^2}{(\delta x)^2}$.  
Eqs.~(\ref{xalpha6})-(\ref{gammaalpha6})  then yield:

\begin{eqnarray}
\label{thetaalpha6}
\frac{\theta_{,N}}{\theta} &=& 
\frac32 (1+w_m) + \frac{\sqrt{3}}{2}
\tilde{\lambda} \theta 
\left[-2q\left(1 + \frac{r}{p}\right)  + \frac{r}{p} \right]
\,,\nonumber\\
\\
\label{betaalha6}
\frac{\beta_{,N}}{\beta} &=& 3 - \sqrt{3} \tilde{\lambda} \theta (1+\beta)\,,
\end{eqnarray}
which has three fixed points given by
\begin{eqnarray}
&&\left(\theta, \beta \right)= \left(0,0\right)\,,\;\;{\rm and}
\label{alpha61}\\
&&
\left(\frac{\sqrt{3}(1+w_m)}
{\tilde{\lambda} \left[2q\left(1+\frac{r}{p}\right)
-\frac{r}{p}\right]},0\right)
\,,\;\;{\rm and}\label{alpha62}\\
&&\left(\frac{\sqrt{3}(1+w_m)}
{\tilde{\lambda} \left[2q\left(1+\frac{r}{p}\right)
-\frac{r}{p}\right]} , \frac{(2q-1) \left(1+\frac{r}{p}\right)
-w_m}{1+w_m}\right)\label{alpha63}\nonumber\\
\end{eqnarray}

For the eigenvalues corresponding to Eq.~(\ref{alpha61}) 
we obtain $3 (1+w_m)/2$ and $3$, which means 
this fixed-point is unstable. 

Considering the fixed-point 
Eq.~(\ref{alpha62}) first, the conditions for $\theta$ to be a finite constant are,
$q>1/2$ or $q<0$. If $0\leq q \leq1/2$, then in the definition of $\theta$, all the terms $\delta x,\, \delta y$ and  $\delta \tilde{\gamma}$ would appear in the numerator, with no apparent singularity present, implying the solution would not be of the form being discussed in this section.  In addition to this, 
as $\delta x$, $\delta y$, $\delta \tilde{\gamma}$ are
all positive, 
$\theta$ must be positive, which gives 
the following condition  
\begin{eqnarray}
&&2q\left(1 +  \frac{r}{p}\right)-\frac{r}{p} >0\,,
\label{exist1alpha6}
\end{eqnarray}
which is $r<-2$ with $p>0$, or $r>-2$ with $p<0$.

Small perturbations around the fixed point Eq.~(\ref{alpha62}),
lead to the following eigenvalues for $\delta \theta$ and $\delta \beta$,
\begin{eqnarray}
w_\theta &=& -\frac{3}{2}(1+w_m)\\
w_\beta &=& 3-\frac{3(1+w_m)}{2q(1+\frac{r}{p})-\frac{r}{p}}
\,,
\end{eqnarray}
where from Eq.~(\ref{exist1alpha6}),
$w_\beta<0$ is satisfied as long as 
\begin{equation}
\label{condition-on-wm}
(1+w_m)>2q(1+\frac{r}{p})-\frac{r}{p}>0,
\end{equation}
which is $-2-p(1+w_m)<r$ with $p>0$, or $-2-p(1+w_m)>r$ 
with $p<0$.

Having shown these new fixed points exist and have well defined behaviour through their combination in Eq.~(\ref{alpha62}), we perturb about them in  Eqs.~(\ref{x_evol_eq_mod}) -(\ref{gamma_evol_eq_mod})
to  obtain the following eignenvalues for the perturbations in $x,\,y$ and $ \tilde{\gamma}$ 
\begin{eqnarray}
w_1 &=& \frac{3}{2} w_m\,,
\label{eigenvalue_xalpha62}\\
w_2 &=& \frac{3}{2}(1+w_m)\left(1-\frac{1}{2q(1+\frac{r}{p})
-\frac{r}{p}}\right)\nonumber\\
&=& \frac{3}{2}(1+w_m)\left(\frac{p+r+2}{r+2}\right)\,,
\label{eigenvalue_yalpha62}\\
w_3 &=& 3\left(1+\frac{\frac{r}{p}(1+w_m)}{2q(1+\frac{r}{p})-
\frac{r}{p}}\right)\nonumber\\
&=&3\left(\frac{2-r w_m}{r+2}\right)\,,
\label{eigenvalue_gammaalpha62}
\end{eqnarray}
where we have used the relation $q=-1/(p+r)$. Although $w_1$, $w_2$ and $w_3$ diverge for $r=-2$,
from Eq.~(\ref{exist1alpha6}), this case is excluded.

Notice that although we have shown eigenvalues for 
3 directions, only two of them are independent
of the $\theta$ direction. This can be seen by the fact that $w_1$, $w_2$ and $w_3$ are related through:
\begin{eqnarray}
(1-2q) w_1 + 2q w_2 +(\frac{1}{2} -q) w_3 =0\,.
\label{rel_eigenvalues_alpha2}
\end{eqnarray}

We can summarise the constraints on the
parameters if the solutions approaching 
the point ($\alpha 6$) given by Eq.~(\ref{alpha62}) are stable. From $q>1/2$ or $q<0$ 
(or equivalently $0>p+r>-2$ or $p+r>0$) and $w_2 < 0$,
we have $r<-2$, which means $-2-p(1+w_m)<r<-2$ with $p>0$ 
from Eqs.~(\ref{exist1alpha6}) and (\ref{condition-on-wm}).
From $w_1 <0$ and $w_3 < 0$
we have $2/r < w_m <0$. 
As $p>0$, $0>1>r/p>-2/p$ for $q>1/2$ and 
$1+r/p >0$ for $q<0$.

We now turn to consider the fixed-point given by (\ref{alpha63}).
In this case only $q>1/2$ is allowed as 
 $\delta x$ and $\delta y$
have the same time dependence. 
Furthermore, as $\beta$ is a nonzero finite constant,
in addition to (\ref{exist1alpha6}),
another condition given by
\begin{eqnarray}
(2q-1)\left(1+\frac{r}{p}\right) -w_m >0\,,
\label{exist2alpha6}
\end{eqnarray}
which is $r<-2-p(1+w_m)$ with $p>0$ or 
$r>-2-p(1+w_m)$ with $p<0$ must be satisfied. 
In fact this extra condition is more restrictive than the one given by 
Eq.~(\ref{exist1alpha6}).

By considering small perturbations around this point,
we obtain the following eigenvalues for $\delta \theta$
and $\delta \beta$,
\begin{eqnarray}
w_\theta &=& -\frac{3}{2}(1+w_m)\\
w_\beta &=& \frac{-3[(2q-1)(1+\frac{r}{p})-w_m]}{2q(1+\frac{r}{p})-\frac{r}{p} } \,,
\end{eqnarray}
which are both negative once the conditions
given by Eqs.~(\ref{exist1alpha6}) 
and (\ref{exist2alpha6}) are satified.

By substituting (\ref{alpha63}) into Eq.~(\ref{x_evol_eq_mod})
we obtain the following eignevalue
for the perturbations in $x$
\begin{eqnarray}
w_1&=& \frac{3 (1+\frac{r}{p})(-1+2q)(1+w_m)}
{2[2q(1+\frac{r}{p})-\frac{r}{p}]}\nonumber\\
&=& -\frac{3(p+r)(2q-1)(1+w_m)}{2(r+2)}\,,
\label{eigenvalue_xalpha63}
\end{eqnarray}
where we have used the relation $q=-1/(p+r)$.

Notice that we have already obtained the stability
for the $\theta$ and $\beta$ directions, the stability based on 
the eigenvalues for the directions of $x$, 
$y$ and $\tilde{\gamma}$ have the same information.
Actually, the eigenvalues corresponding to
the perturbation in $y$ and $\tilde{\gamma}$ are given by
\begin{eqnarray}
w_2&=& \frac{3 (1+\frac{r}{p})(-1+2q)(1+w_m)}
{2[2q(1+\frac{r}{p})-\frac{r}{p}]}\,,\\
w_3&=& \frac{6 (1+\frac{r}{p})(1+w_m)}
{2[2q(1+\frac{r}{p})-\frac{r}{p}]}\\
&=& -\frac{3(p+r)(1+w_m)}{(r+2)}
\,,
\end{eqnarray}
which are consistent with
\begin{eqnarray}
&&w_1=w_2\,,\\
&&(-2q+1) w_1 + 2q w_2 + (\frac12 - q) w_3=0\,,
\end{eqnarray}
derived from the fact that $\theta$ and $\beta$ are constant.
Although $w_1$, $w_2$ and $w_3$ diverge for $r=-2$,
from Eq.~(\ref{exist1alpha6}), this case is excluded.

We can summarise the constraints on the
parameters if the solutions approaching 
the point ($\alpha 6$) given by Eq.~(\ref{alpha63}) are stable. From $q>1/2$
(or equivalently $0>p+r>-2$) and $w_{1,2,3} < 0$,
we have $r<-2$, which means $-2-p(1+w_m)>r$ with $p>0$ 
from Eq.~(\ref{exist2alpha6}). 
If we interpret this as the constraint on $w_m$,
together with $q>1/2$, $w_m <0$ should be also
satisfied. As $p>0$, $0>1+r/p>-2/p$ for $q>1/2$.

It is worth mentioning that for the fluid dominated 
solution $(\alpha 6)$
realised both through Eqs.~(\ref{alpha62}) and (\ref{alpha63}), 
$w_m < 0$ is required. Such a restriction implies that we can not be considering the usual matter or radiation fluid as the background for $(\alpha 6)$ to be stable under, although it is consistent for example with that of a cosmological constant.

We finish this subsection by providing a summary of the solutions $(\alpha1)$ to $(\alpha6)$ and their stability criteria in Table~\ref{fixed-points-Summary_0/0} below.

\begin{widetext}
\begin{center}
\begin{table} [h!]
\begin{tabular} { |c|c|c|c|c|c|c| }
\hline 
 & $x$ & $y$ & $\tilde{\gamma}$ & $\Omega_\phi$ & Valid $q$ & 
Stability \\
\hline
$(\alpha1)$ & $ 0$ & $1$ & $1$ &$1$ &$q<-1/2$ and $q>0$ &  
marginally stable\\      
\hline
$(\alpha2)$ & $0$ & $1$ & $0$ & $1$ & $-1/2 < q < 1/2$ &  
marginally stable\\ 
\hline
$(\alpha3)$ & $1$ & $0$ & $0$ & $1$ & $q < 0$ & $w_m >
 0,\;\;$ $r<-2,\;\;$$p>0,\;\;$  $\frac{r}{p} >-1$ \\    
& & & & & $q >1/2 $& $w_m >
 0,\;\;$$r<-2,\;\;$$p>0,\;\;$$-1-\frac{2}{p} < \frac{r}{p} <-1$  \\   
\hline
$(\alpha4)$ & $\;x_0\;$$\;(0<x_0 < 1)$ & $0$ & $0$ & $x_0 ^2$ & $q < 0$
 & $w_m =0,\;\;$ $r<-2,\;\;$$p>0,\;\;$  $\frac{r}{p} >-1$
\\   
& & & & & $q >1/2 $& $w_m =
 0,\;\;$$r<-2,\;\;$
$p>0,\;\;$$-1-\frac{2}{p} < \frac{r}{p} <-1$  \\     
\hline
$(\alpha5)$ & $0$ & $0 $ & 
$1$ & $0$ & $q<0$ & 
$\frac{p-6}{p+2}>w_m>-1,\;\;$$p>2,\;\;$
 $\frac{r}{p} >  -1$ \\
\hline
$(\alpha6)$ & $ 0$ & $0$ & $0$ &$0$ &$q < 0$ & 
$0>w_m >\frac{2}{r},\;\;$$p>0,\;\;$$-2>r>-2-p(1+w_m),\;\;$
$\frac{r}{p}>-1$ \\   
& & & & & $q >1/2 $& $0>w_m(>\frac{2}{r}),\;\;$
$p>0,\;\;$ $-2>r,\;\;$ $-1-\frac{2}{p} < \frac{r}{p} <-1$
\\   
\hline

\end{tabular}
\caption{This table summarises the validity and stability of the fixed-points $\alpha1$ to $\alpha6$. Notice that we have restricted ourselves to the regime 
$\tilde{\lambda} > 0$ and $x \geq 0$. For the condition for
stability of $(\alpha 6)$ for $q>1/2$, $w_m > 2/r$
holds only for $r>-2-p(1+w_m)$.
}
\label{fixed-points-Summary_0/0}
\end{table}
\end{center}
\end{widetext}

\subsection{Late-time behaviour}
\label{subsec_pow_latetime}

Now that we have analyzed the stability of the fixed-point
solutions, we are in a position to discuss the late-time
attractor structure of the autonomous system described by
Eqs.~(\ref{x_evol_eq_mod})-(\ref{gamma_evol_eq_mod}).
It is clear from the solutions presented in
TABLE~\ref{fixed-points-Summary_power} and 
\ref{fixed-points-Summary_0/0}
that the precise structure of the solutions to the 
dynamical system depends on the value of $q$. 
 We have seen that for the particular cases of 
$q=1/2$, $q=0$, $q=-1/2$
there are non-trivial ``scaling solutions'' 
where $x$, $y$ and $\tilde{\gamma}$ 
are finite constants depending on the model parameters
$\tilde{\lambda}$, $\tilde{\mu}$, and the equation of state
of the fluid $w_m$.
Therefore, in this subsection, first we will discuss
the late-time attractor structure of the system
with special emphasis on these three cases,
and then we will move on to a discussion of
the other cases.

\subsubsection{The case with $q=1/2$}

TABLE~\ref{fixed-points-Summary_power},
shows that in the case of $\tilde{\lambda} + \tilde{\mu} <0$
(or equivalently $p>0$), 
if $w_m \geq 0$, the standard fixed-point $(a4)$
(ultra-relativistic kinetic-potential scaling solutions)
is stable for $w_m > -1 + \tilde{\lambda} 
(\sqrt{\tilde{\lambda}^2-12} - \tilde{\lambda} )/6$.
 It is also stable
for $w_m < 0$ if  
$\tilde{\lambda} < \sqrt{3}(1+w_m)/\sqrt{-w_m}$, which also corresponds to the requirement that $p>0$ with $p+r=-2$.
On the other hand, if $\tilde{\lambda} \geq
\sqrt{3}(1+w_m)/\sqrt{-w_m}$ with $w_m<0$ (or equivalently
$p \geq \sqrt{3 \sigma \nu}(1+w_m)/\sqrt{-w_m} >0$),
then the standard fixed-point $(a5)$
(ultra-relativistic kinetic-potential-fluid scaling solution)
is the stable solution. Clearly, these two fixed-point solutions
are candidates for the late-time attractor solution to the system 
as long as the model parameters satisfy the conditions mentioned above. 
On the other hand, TABLE~\ref{fixed-points-Summary_0/0},
shows that for $q=1/2$, the fixed-point associated with $(\alpha 1)$ 
(standard potential dominated solutions)
is marginally stable.  As $(\alpha 1)$ has a $zero$ 
eigenvalue, to specify which of the three possibilities 
$(a4)$, $(a5)$ or ($\alpha 1$) is likely to win out will require that we go
to higher order, although it is likely to be $(a4)$ or $(a5)$ as one of
these is definitely stable when the conditions
$\tilde{\lambda}+\tilde{\mu}<0$ and $q=1/2$ are satisfied. 
It is worth mentioning that for the cosmologically
sensible cases, we expect $w_m \geq 0$. Then, only 
$(a4)$ can be the late-time attractor.

\begin{center}
\begin{figure}[t]
\centering
 \scalebox{.8}
 {\rotatebox{0}{
    \includegraphics*{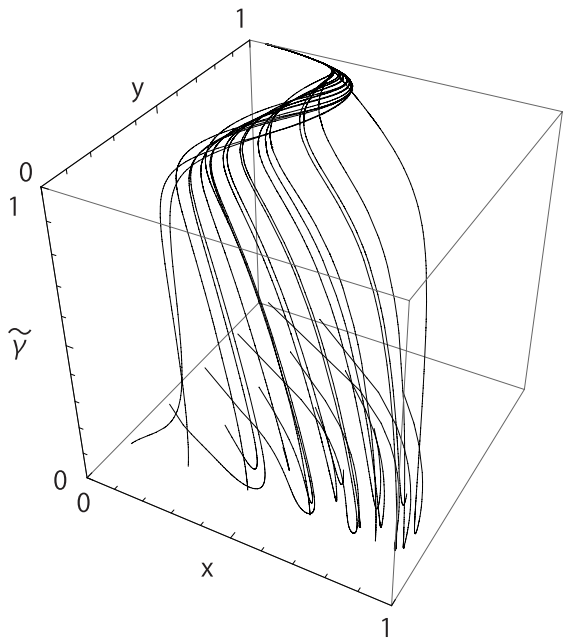}
                 }
 }
\begin{widetext}
\caption{Demonstration of the attractive property of $(\alpha 1)$ in the
 regime $\tilde{\lambda}+\tilde{\mu}>0$ and $q=1/2$.
Note that for all the initial values of $x$, $y$ and $\tilde{\gamma}$,
the solutions approach that of $\alpha 1$ where  $(x,y,\tilde{\gamma}) = (0,1,1)$.
Note also that the physical region is confied to $x^2+y^2 \leq 1$
in the three-dimensional phase space.
Although here we have chosen   $q=1/2$, $r/p=-0.5$, 
$\tilde{\lambda}=10$ (or $p= -4$, $r=2$, $(\sigma \nu)^{1/2} 
=2/5$), $w_m = 0$, the same results are obtained for $q=1/2$
with $r/p >-1$. }
\label{numerical_alpha1_2}
\end{widetext}
\end{figure}
\end{center}

Alternatively, for $\tilde{\lambda}+\tilde{\mu}>0$ 
(or equivalently $p<0$), as both points 
$(a4)$ and $(a5)$ have positive eigenvalues,
we can expect $(\alpha 1)$ to be the late-time attractor. 
These findings have been  confirmed numerically as can be seen in 
Fig.~\ref{numerical_alpha1_2}. 

As we have already mentioned, $(a4)$ and $(a5)$ are 
new solutions found only in the case of the DBI field.
$(a4)$ corresponds to the ultrarelativistic 
version of the usual canonical power-law inflationary solution. It is well known that power-law inflation is strongly
constrained by the observation of the spectral index of primordial
perturbations (for example, see \cite{Komatsu:2010fb}).
Although it is difficult to have 
DBI driven power-law inflation,
we can in principle use it to explain the current acceleration of the universe.
On the other hand, although $(a5)$ is also 
unique to the DBI field, because the solution 
requires $w_m < 0$ to be satisfied, it is not so realistic given that we are usually considering $w_m$ to be either matter or radiation.

However, since  $(\alpha 1)$ corresponds to the usual accelerating solution
where the scalar field rolls slowly down its potential and is not unique to
the DBI field, this solution can in principle be used to explain both 
an early stage of inflation and the present dark energy dominated period of acceleration.

\begin{center}
\begin{figure}[t]
\centering
 \scalebox{.8}
 {\rotatebox{0}{
    \includegraphics*{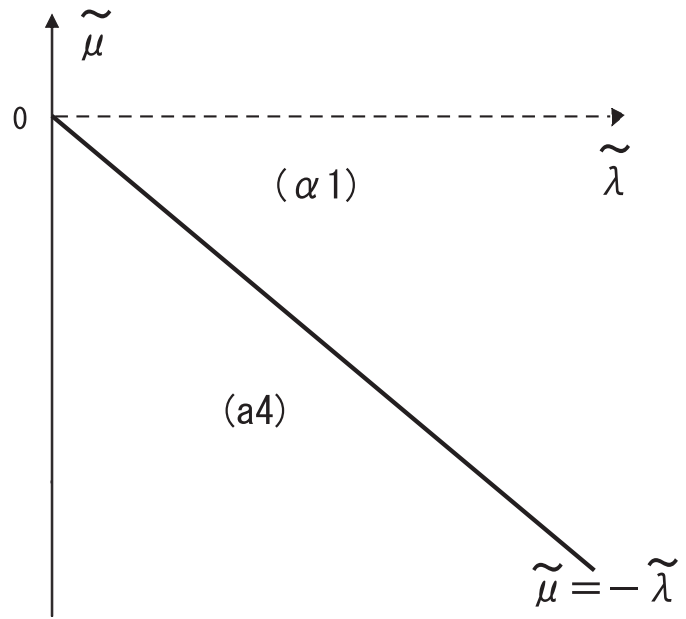}
                 }
 }
\begin{widetext}
\caption{Late-time attractor structure in the models 
where the potential and the brane tension are 
power-law functions of the DBI field with $q=1/2$
and $w_m \geq 0$. 
We show the late-time attractor solution for a given model
parameter in 
$\tilde{\lambda}$ - $\tilde{\mu}$ space with 
$\tilde{\lambda}>0$.  In the case with $w_m < 0$, the region where $(a5)$ is the late-time attractor appears for $\tilde{\lambda} + 
\tilde{\mu} < 0$ and $\tilde{\lambda} \geq 
\sqrt{3} (1+w_m)/\sqrt{-w_m}$.
It is worth noting that $\tilde{\lambda}$ and 
$\tilde{\mu}$ are related with the parameters
of the potential and brane tention as
$\tilde{\lambda} = |p|/\sqrt{\sigma \nu}$
and $\tilde{\mu} = -(2+p) |p|/(p \sqrt{\sigma \nu})$.}
\label{fig_half_attractor}
\end{widetext}
\end{figure}
\end{center}

In Fig.~\ref{fig_half_attractor}, we summarise 
the late-time attractor structure with $q=1/2$ and 
$w_m \geq 0$.

\subsubsection{The case with $q=0$}

Here, we consider the late-time attractor structure
in the models where the potential and the brane tension
are power-law functions of the DBI field with $q=0$.

As is shown in TABLE~\ref{fixed-points-Summary_power},
in the case with $\tilde{\lambda} + \tilde{\mu} >0$
(or equivalently $1+r/p>0$), 
if $\tilde{\lambda} < \sqrt{3(1+w_m)}$
(or equivalently $|p| < \sqrt{3 (1+w_m)}$), 
the fixed-point $(b2)$
(the standard kinetic-potential scaling solution)
is  stable.
If $\tilde{\lambda} \geq \sqrt{3(1+w_m)}$
(or equivalently $|p| \geq \sqrt{3 (1+w_m)}$),
the fixed-point $(b3)$
(standard kinetic-potential-fluid scaling solutions)
is stable.
Furthermore, if the conditions 
$\tilde{\mu} < -\sqrt{6}$ and $w_m > 3/(\tilde{\mu}^2-3)$ 
(or equivalently $w_m > 3/(r^2-3)$) are satisfied,
$(c1)$
(the relativistic kinetic energy dominated solution)
is stable.
Therefore, these fixed-point solutions
are clearly candidates for the late-time attractor behaviour, provided
the model parameters satisfy the conditions mentioned above.
We also find for the region in the parameter space
satisfying both stability conditions for $(b3)$ and $(c1)$,
which is the late-time attractor depends on the initial 
values of $x$, $y$ and $\tilde{\gamma}$. 
On the other hand, TABLE~\ref{fixed-points-Summary_0/0},
shows that for $q=0$, the fixed-point $(\alpha 2)$ (ultrarelativistic potential dominated solutions)
is marginally stable.  

Using arguments similar to those applied for the case with $q=1/2$,
we expect that for $\tilde{\lambda}+\tilde{\mu}>0$ 
(or equivalently $1+r/p>0$), 
one of  $(b2)$, $(b3)$, $(c1)$ will be the late-time
attractor solution of this system as long as their stability conditions are satisfied. 
This is because we have seen that the stability properties 
of these three points are given by three
negative eigenvalues which is a stronger condition than for $(\alpha 2)$.

\begin{center}
\begin{figure}[t]
\centering
 \scalebox{.8}
 {\rotatebox{0}{
    \includegraphics*{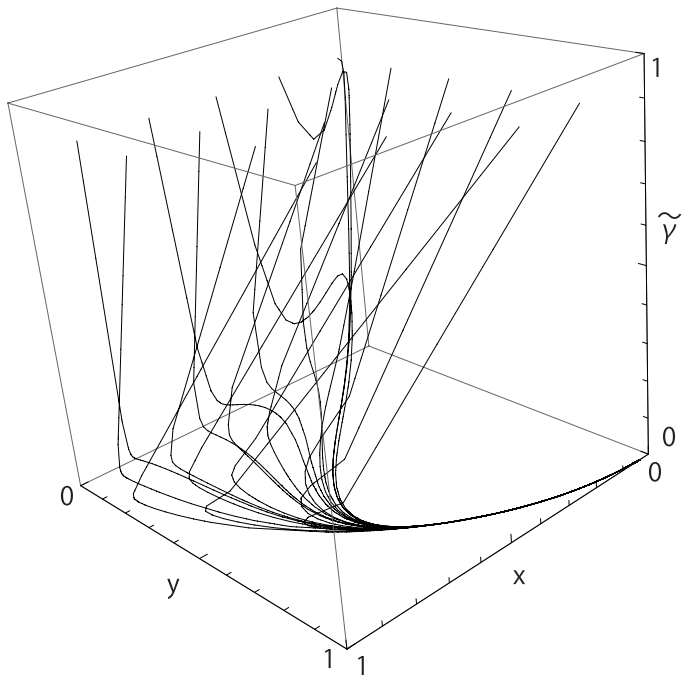}
                 }
 }
\begin{widetext}
\caption{Demonstration of the attractive property of $(\alpha 2)$ in the
 regime $\tilde{\lambda}+\tilde{\mu}<0$ and $q=0$.
Note that for all the initial values of $x$, $y$ and $\tilde{\gamma}$,
the solutions approach that of $(\alpha 2)$ where  $(x,y,\tilde{\gamma}) = (0,1,0)$.
Note also that the physical region is confied to $x^2+y^2 \leq 1$
in the three-dimensional phase space.
Although here we have chosen   $q=0$, $r/p=-1.5$, 
$\tilde{\lambda}=10$, $w_m = 0$, 
the same results are obtained for $q=0$
with $r/p < -1$. 
Notice also  that 
we chose the directions of the axis differently
from Fig.~\ref{numerical_alpha1_2}}
\label{numerical_alpha2_0}
\end{widetext}
\end{figure}
\end{center}

On the other hand for $\tilde{\lambda}+\tilde{\mu}<0$ 
(or equivalently $1+r/p<0$), the points 
$(b2)$, $(b3)$ and $(c1)$ have positive eigenvalues,
so we expect $(\alpha 2)$ will become the late-time attractor. 
This has once again been confirmed numerically in  
Fig.~\ref{numerical_alpha2_0}.

Since $(b2)$ and $(b3)$ correspond to the well known
power-law inflationary and scaling solutions
respectively, the cosmology based on these solutions
has already been well studied
\cite{Lucchin:1984yf,Ratra_Peebles,trac,Ferreira:1997hj,Copeland:1997et}. As in the ultra-relativistic case, the power-law inflationary
solution $(b2)$ can explain the present acceleration, while
the scaling solution $(b3)$ 
can play a very important role in classifying the late-time attractor structure of the system. 

As we have mentioned earlier, $(c1)$ is a new solution specific
to the DBI field, but we do not consider it further because it can neither explain the current acceleration of the Universe, nor can it accommodate a fluid. 

$(\alpha 2)$ is also a solution specific to the DBI field
and is in fact a very interesting new type of inflationary solution. 
These models are however tightly constrained. They have been shown to 
give too large a degree of primordial non-Gaussianity \cite{Silverstein:2003hf,Alishahiha:2004eh,Chen:2004gc,Chen:2005ad,Shandera:2006ax}.

In Fig.~\ref{fig_zero_attractor}, we summarise 
the late-time attractor structure with $q=0$ and $w_m > 0$.

\begin{widetext}
\begin{center}
\begin{figure}[t]
\centering
 \scalebox{.8}
 {\rotatebox{0}{
    \includegraphics*{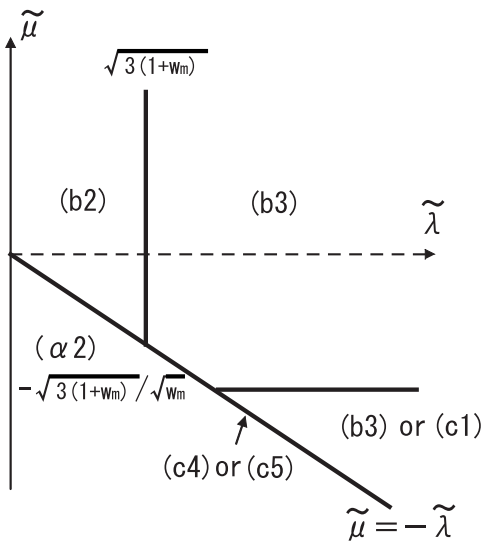}
                 }
 }
\caption{Late-time attractor structure 
in the models where the potential and brane tension
are power-law functions of the DBI field with $q=0$
and $w_m >0$. 
We show which fixed-point solution
is the late-time attractor solution for a given set of model
parameters in  $\tilde{\lambda}$-$\tilde{\mu}$ 
space with $\tilde{\lambda}>0$.
In the case with $w_m \leq 0$, the region where $(c1)$
is the late-time attractor disappears.
This figure can be also applied to the models 
where the potential and brane tension are 
exponential function of the DBI field with
$q=0$ through the
replacements $\tilde{\lambda} (=|p|) \to \lambda$, 
$\tilde{\mu} (=r |p| /p )\to \mu$.
For this class of models, it is possible to consider
the case with $\mu=-\lambda$ where $(c4)$ or $(c5)$
can be the late-time attractor.
Although it is complicated to write down, there is a critical
value of $\lambda$ depending on $w_m$, $\sigma$ and $\nu$,  
below which $(c4)$ is the late-time attractor
and above which $(c5)$ is the late-time attractor. 
}
\label{fig_zero_attractor}
\end{figure}
\end{center}
\end{widetext}

\subsubsection{The case with $q=-1/2$}

Here, unlike the previous cases with $q=1/2$
and $q=0$, when $q=-1/2$, the stable solutions with three negative
eigenvalues for $\tilde{\lambda} + \tilde{\mu} >0$
(or equivalently $p>0$), are given by the four fixed-points ($(\alpha 3)$ - $(\alpha 6)$) shown in 
TABLE~\ref{fixed-points-Summary_0/0}.  

$(\alpha 3)$ (ultrarelativistic kinetic dominated solutions)
is stable for $w_m \geq 0$ with $p>4$, while
$(\alpha 4)$ (Ultrarelativistic kinetic-fluid scaling solutions) is
stable for $w_m =0$ with $p>4$.
$(\alpha 5)$ (Standard fluid dominated solutions)
is stable for $(p-6)/(p+2) > w_m > -1,\;\;$ with 
$p>2$, and $(\alpha 6)$ (Ultrarelativistic fluid dominated solutions)
is stable for $0>w_m > 2/(2-p),\;\;$ with 
$4/(-w_m) > p > 4$. 

These four fixed-points
are candidates  to be the late-time attractor solutions
to the system as long as the model parameters satisfy
the conditions mentioned above. 
When a region of parameter space allows more than one stable fixed-point, which one of them becomes 
the late-time attractor depends on the initial values of $x$, $y$ and $\tilde{\gamma}$. 
On the other hand, from TABLE~\ref{fixed-points-Summary_power},
we see that in this case the fixed-point $(c3)$ 
(the relativistic potential dominated solution) is marginally
stable. Therefore $(c3)$ is also 
a candidate late-time attractor solution. However, as we saw before, when a marginally stable solution also exists, we expect that for $\tilde{\lambda}+\tilde{\mu}>0$ 
(or equivalently $p>0$),  one of the four fixed points  $(\alpha 3)$-$(\alpha 6)$ 
will be the late-time  attractor solution of the system 
in the region of the phase space  when their stability conditions are satisfied.

\begin{center}
\begin{figure}[t]
\centering
 \scalebox{.8}
 {\rotatebox{0}{
    \includegraphics*{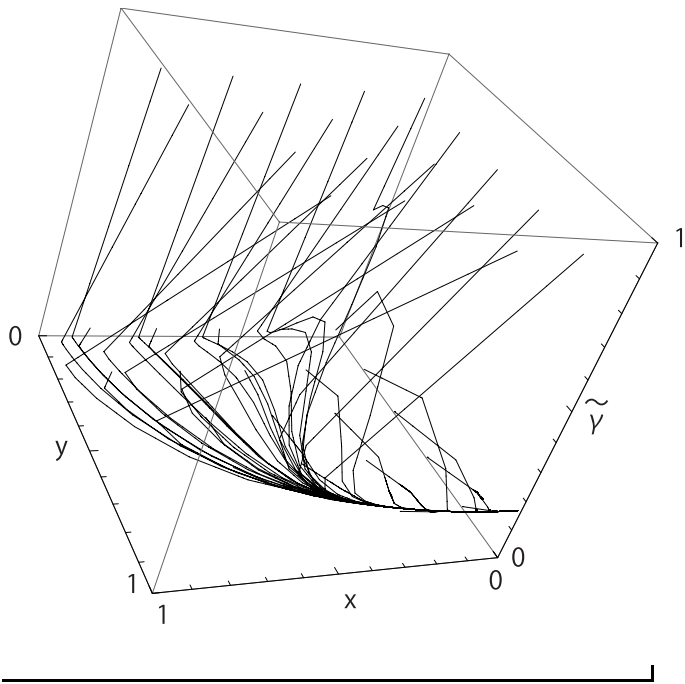}
                 }
 }
\begin{widetext}
\caption{Demonstration of the attractive property of $(c 3)$ in the
 regime $\tilde{\lambda}+\tilde{\mu}<0$ and $q=-1/2$.
For every initial values of $x$, $y$ and $\tilde{\gamma}$,
the solution approaches $(x,y,\tilde{\gamma}) = 
(0,1,\sqrt{3}/\sqrt{\tilde{\lambda}^2+3})$.
Notice that the physical region is confined to $x^2+y^2 \leq 1$
in the three-dimensional phase space.
Although we choose   $q=-1/2$, $r/p=-1.5$, $\tilde{\lambda}=10$
(or $p=-4$, $r=2$, $(\sigma \nu)^{1/2} =2/5$) and $w_m=0$
in this example, the same results are obtained for 
$q=-1/2$  with $-1 > r/p$. Notice that
for $\tilde{\lambda} =10$, 
$\sqrt{3}/\sqrt{\tilde{\lambda}^2+3} \simeq 0.17$ as expected.}
\label{numerical_c3}
\end{widetext}
\end{figure}
\end{center}

Turning our attention instead to  consider the regime $\tilde{\lambda}+\tilde{\mu}<0$ 
(or equivalently $p<0$), we see that the fixed-points 
$(\alpha 3)$ - $(\alpha 6)$  now have positive eigenvalues,
and so we expect $(c3)$ to become the late-time attractor, a result that we have confirmed numerically in 
Fig.~\ref{numerical_c3}.

Cosmology based on  the existence of the solutions $(\alpha 3)$ and $(\alpha 4)$ 
are peculiar to the DBI field in the sense that 
the field behaves like dust even though the kinetic term completely
dominates the potential term. However, although interesting in its own
right, the solutions do not appear to be useful cosmologically. 
Solutions  $(\alpha 5)$ or  $(\alpha 6)$ both imply that 
the Universe is completely dominated by the background fluid
with the DBI field playing a negligible role in its evolution. 

On the other hand, the solution $(c3)$ appears to have some interesting features 
specific to the DBI field. In this case, $\tilde{\gamma}$
or equivalently  the sound speed $c_s$ can be a constant ranging between $0$ and $1$.
In such a case, if this solution is realised in the very early 
Universe, it opens up the possibility of a large but still
allowed degree of non-Gaussianity being obtained.
Furthermore, since the current prediction of non-Gaussianity
from this type of inflation model is based on the assumption
that $c_s$ is constant, this new solution
serves as a good concrete background about which to consider 
cosmological perturbations.

\subsubsection{The case with general $q \neq 0,\,1/2$ or $-1/2$}

To complete the classification of the
late-time attractor structure in the models where
the potential and the brane tension are power-law
functions of the DBI field, we now consider
the case where  $q$ is different from $1/2$, $0$ or $-1/2$. For three cases 
we have earlier seen that the question of whether marginally stable solutions 
can be the late-time attractor or not depends 
upon the stability of the other allowed fixed-points.
In particular those fixed points with three negative eigenvalues, means that these marginally stable
fixed-point solutions can not be the late-time attractor
as their stability is weaker than that of the fixed-points. However, 
if all other fixed-points have  positive
eigenvalues, then the marginally stable fixed-point
solutions can turn out to be the late-time attractor. 

Significantly, since this criteria also holds for the cases considered here,
we will apply it without showing the numerical plots except the special case
with $0<q<1/2$. 

For $q>1/2$, for $\tilde{\lambda} + \tilde{\mu} < 0$
(or equivalently $p >0$), 
there are four stable fixed-point solutions
$(\alpha 3)-(\alpha 6)$. In addition to these, the fixed-point solution
$(\alpha 1)$ is marginally stable.

As in the previous cases, for 
$\tilde{\lambda} + \tilde{\mu} < 0$,
one of the  fixed-point solutions
$(\alpha 3)-(\alpha 6)$ can be the late-time attractor
depending on the values of parameters $p$, $r$, $w_m$
or the initial values of $x$, $y$, $\tilde{\gamma}$,
while for $\tilde{\lambda} + \tilde{\mu} > 0$
(or equivalently $p < 0$), $(\alpha 1)$ is the 
late-time attractor. 

\begin{center}
\begin{figure}[t]
\centering
 \scalebox{.8}
 {\rotatebox{0}{
    \includegraphics*{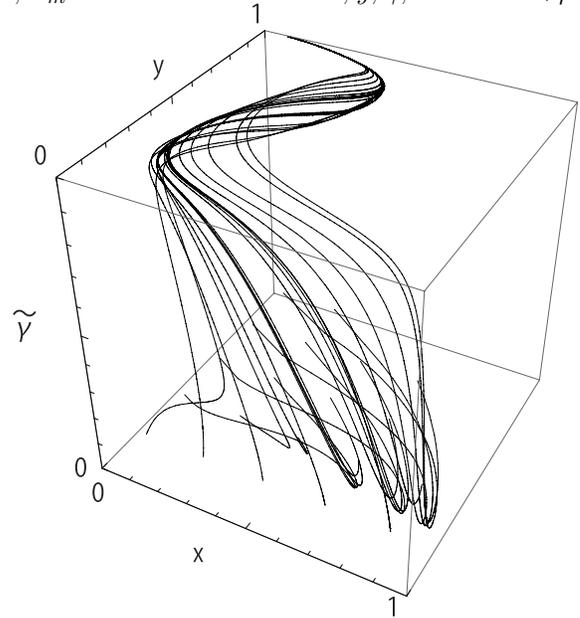}
                 }
 }
\begin{widetext}
\caption{Demonstration of the attractive property of $(\alpha 1)$ in the regime $\tilde{\lambda}+\tilde{\mu}>0$.
Note that for all the initial values of $x$, $y$ and $\tilde{\gamma}$,
the solutions approach that of $\alpha 1$ where  $(x,y,\tilde{\gamma}) = (0,1,1)$.
Note also that the physical region is confied to $x^2+y^2 \leq 1$
in the three-dimensional phase space.
Although here we have chosen   $q=1/4$, $r/p=-0.5$, 
$\tilde{\lambda}=10$ (or $p=-8$, $r=4$, 
$(\sigma \nu)^{1/4} =4/5$),$w_m=0$, the same results are obtained for $q>0$ 
with $r/p > -1$  and $-1/2 > q$ with $r/p < -1$.}
\label{numerical_alpha1}
\end{widetext}
\end{figure}
\end{center}

\begin{center}
\begin{figure}[t]
\centering
 \scalebox{.8}
 {\rotatebox{0}{
    \includegraphics*{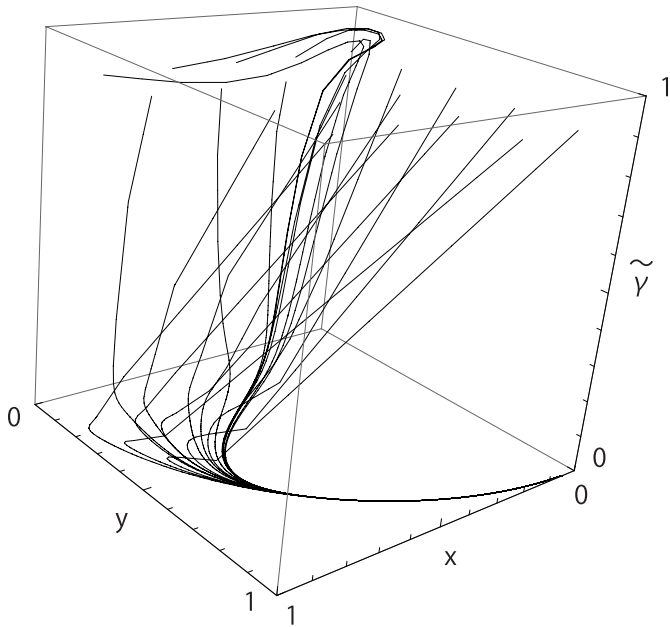}
                 }
 }
\begin{widetext}
\caption{Figure showing the attractive nature of the fixed point  $(\alpha 2)$
in the regime $\tilde{\lambda}+\tilde{\mu}<0$ and $q=\frac14$. 
For all initial values of $x$, $y$ and $\tilde{\gamma}$,
solutions approach the point $(\alpha 2)$ where $(x,y,\tilde{\gamma}) = (0,1,0)$.
Notice that physical region is confined to $x^2+y^2 \leq 1$
in the three-dimensional phase space.
Although we have chosen  $q=1/4$, $r/p=-1.5$, $\tilde{\lambda}=10$
(or $p=8$, $r=-12$, $(\sigma \nu)^{1/4} =4/5$),
$w_m=0$ in this example, the same results are obtained for 
$1/2>q>-1/2$  with $-1 > r/p$. Notice also  that 
we chose the directions of the axis differently
from Fig.~\ref{numerical_alpha1_2}}
\label{numerical_alpha2}
\end{widetext}
\end{figure}
\end{center}

For $0<q<1/2$, there are two marginally stable fixed-point
solutions ($\alpha1$) and ($\alpha2$).
As there are no other fixed-points whose stability 
around the points are characterised by three negative eigenvalues,
determining which of the two will be the late-time attractor is non-trivial.
However, numerically we find for 
$\tilde{\lambda} + \tilde{\mu} > 0$ (or equivalently $p <0$),
$(\alpha 1)$ is the late-time attractor whereas
for $\tilde{\lambda} + \tilde{\mu} < 0$ 
(or equivalently $p >0$), it is $(\alpha 2)$.
(See Figs.~\ref{numerical_alpha1} and \ref{numerical_alpha2})

In the case where $-1/2<q<0$, when $\tilde{\lambda} + \tilde{\mu} > 0$
(or equivalently $p > 0$), 
there are four stable fixed-point solutions
$(\alpha 3)-(\alpha 6)$, with the fixed-point solution
$(\alpha 2)$ being marginally stable.
One of $(\alpha 3)-(\alpha 6)$ can be the late-time attractor
depending on the values of the parameters $p$, $r$, $w_m$
or the initial values of $x$, $y$, $\tilde{\gamma}$,
while for $\tilde{\lambda} + \tilde{\mu} < 0$
(or equivalently $p < 0$), $(\alpha 2)$ is the 
late-time attractor. 

When $q< -1/2$, for  $\tilde{\lambda} + \tilde{\mu} > 0$
(or equivalently $p > 0$), the story is same as in the 
case with $0 > q > -1/2$,
there are four stable fixed-point solutions
$(\alpha 3)-(\alpha 6)$. But in this case,
an additional marginally stable solution is $(\alpha 1)$. Once again, one of $(\alpha 3)-(\alpha 6)$ can be the late-time attractor solutions
depending on the values of parameters $p$, $r$, $w_m$
or the initial values of $x$, $y$, $\tilde{\gamma}$,
while for $\tilde{\lambda} + \tilde{\mu} < 0$
(or equivalently $p < 0$), $(\alpha 1)$ is the 
late-time attractor.

We summarise  the possible late-time attractor solutions
for a given set of $q$, $p$, $r$ 
based on the discussion in this subsection in  
TABLE~~\ref{attractor-Summary_power}.

\begin{widetext}
\begin{center}
\begin{table} [h!]
\begin{tabular} { |c|c|c| }
\hline 
 & $1+\frac{r}{p} >0$&
$1+\frac{r}{p} <0$
 \\
\hline
$q>\frac{1}{2}$ & $(\alpha 1)$&$(\alpha 3)$ or 
$(\alpha 4)$ or $(\alpha 5)$ or $(\alpha 6)$
 \\
\hline
$q=\frac{1}{2}$ & $(\alpha 1)$&$(a4)$ or $(a5)$
 \\
\hline
$\frac{1}{2} > q > 0$ & $(\alpha 1)$&$(\alpha 2)$ 
 \\
\hline
$q=0$ & $(b2)$ or $(b3)$ or $(c1)$ &$(\alpha 2)$ 
 \\
\hline
$0>q>-\frac{1}{2}$ & $(\alpha 3)$ or 
$(\alpha 4)$ or $(\alpha 5)$ or $(\alpha 6)$
&$(\alpha 2)$ 
 \\
\hline
$q =-\frac{1}{2}$ & $(\alpha 3)$ or 
$(\alpha 4)$ or $(\alpha 5)$ or $(\alpha 6)$ 
&$(c3)$ 
 \\
\hline
$-\frac{1}{2} > q$ & $(\alpha 3)$ or 
$(\alpha 4)$ or $(\alpha 5)$ or $(\alpha 6)$ &
$(\alpha 1)$
 \\
\hline
\end{tabular}
\caption{This table summarises which fixed-points
can be the late-time attractors for the models 
where the potential and brane-tension are power-law
functions of the DBI field for a given 
$q$, $p$ and $r$.  
``or'' means there may be more than one possible late-time
attractor and which is realised depends on $w_m$
and the initial value of $x$, $y$, $\tilde{\gamma}$, 
which has been described in 
TABLE~\ref{fixed-points-Summary_power} and 
TABLE~\ref{fixed-points-Summary_0/0} in detail.
}
\label{attractor-Summary_power}
\end{table}
\end{center}
\end{widetext}


\section{Models with exponential potential and brane tension
\label{section_exp}}

In this section
we consider the models where both the potential 
and brane tension are exponential functions
of the DBI field,

\begin{eqnarray}
V(\phi) = \sigma e^{-\lambda \phi},
\;\;\;\;\;\;\;\;\;\;\;\;
f(\phi) = \nu e^{-\mu \phi},
\label{lambda_mu_exp_func}
\end{eqnarray}
where $\sigma$ and $\nu$ are constants. We will see that there is a direct link with the $q=0$ limit for the scaling solutions (b2), Eq.~(\ref{b2}) and(b3), Eq.~(\ref{b3}) obtained in the previous section.

\subsection{Autonomous System}

In this case, from the discussions in Sec.~\ref{basic-eqn},
in terms of constants $\lambda$ and $\mu$ given by 
Eq.~(\ref{lambda_mu_exp_func}),
Eqs.~(\ref{x_evol_eq_exp}) - (\ref{gamma_evol_eq_exp})
constitute an autonomous system.

These equations correspond exactly to
Eqs.~(\ref{x_evol_eq_mod})-(\ref{gamma_evol_eq_mod}) with $q=0$ under
the identification $\lambda \to \tilde{\lambda} (= |p|)$ 
and $\mu \to \tilde{\mu}( = r |p| /p)$. 
It is as expected because the exponential function can be regarded as the power law function $\phi^p$ with $p \to \infty$. This means that all the fixed points associated with $q=0$ are found in the case of the exponential functions as well. However, as we shall now show, when $\lambda + \mu=0$, there is an additional fixed point found in the exponential case.
As in the previous section, without loss of generality, we restrict our discussion to the case $\lambda >0$ and $x \geq 0$.

\subsection{Fixed-point Solutions}
\label{fixed-point-exp-pot}

From our discussion in the power-law case for the 
potential and brane tention with $q=0$, the fixed-point
solutions obtained there 
are also fixed point solutions for the exponential case. 
This means that the eight fixed-point solutions
$(a1),\,(a2),\,(b1)-(b3),\,(c1),\,(c2), (\alpha 2)$ 
obtained in Sec.~\ref{section_powerlaw} are also 
fixed-point solutions with the replacement 
$\tilde{\lambda} (= |p|) \to \lambda$
and $\tilde{\mu} (=r |p|/p) \to \mu$. 

The additional solution not present in the power law case follows when
\begin{eqnarray}
\mu = -\lambda\,.
\label{rel_mu_lambda_yneq0} 
\end{eqnarray}
As we will now show a class of new solutions then exist for 
$0\leq \tilde{\gamma} \leq 1$ and non-zero $x$ and $y$ .

Strictly speaking, since in this case $V(\phi)$ and $f(\phi)$
are related through the constraint $f V = \sigma \nu$, this is a two-dimensional system
as opposed to a three-dimensional one.
Actually $\tilde{\gamma}$ can be completely specified
once $x$ and $y$ are given, through the relation
\begin{eqnarray}
\tilde{\gamma} = \frac{y^2}{3 \sigma \nu x^2 + y^2}\,.
\label{const_gamma_ito_xy_ex_sp}
\end{eqnarray}
However, for simplicity we continue to solve explicitly 
the evolution equation of $\tilde{\gamma}$
to obtain the fixed-points.
Of course, from Eq.~(\ref{const_gamma_ito_xy_ex_sp}),
$\tilde{\gamma}$ becomes constant when $x$ and $y$
become constant, which provides a consistency check for the method. In fact the constraint
Eq.~(\ref{const_gamma_ito_xy_ex_sp}),
leads to some more interesting constraints on the DBI field. Substituting it 
into Eq.~(\ref{w_phi}) and defining $r \equiv y^2/x^2$,
the equation of state of the DBI field is given by
\begin{eqnarray}
w_\phi (r) = \frac{r(1-3 \sigma \nu -r)}{(3 \sigma \nu +r)
(1+r)}\,,\;\;\;\;{\rm with}\;\;0 \leq r\,,
\end{eqnarray}
which clearly asymptotes to $0$ as $r \to 0$, 
while $-1$ as $r \to \infty$.
In fact the  function decreases monotonically 
and is constrained as 
\begin{eqnarray}
-1 \leq w_\phi < 0\,.
\label{w_phi_sp_exp}
\end{eqnarray}

To find the fixed-points in the system, substituting Eq.~(\ref{rel_mu_lambda_yneq0}) into Eq.~(\ref{gamma_evol_eq_exp}), and requiring $\tilde{\gamma}$ remain constant implies,
\begin{eqnarray}
1-x^2-y^2=1-
\frac{\sqrt{3(1+\tilde{\gamma})}}
{\sqrt{\tilde{\gamma}}\lambda} x.
\label{rel_x_y_yneq0}
\end{eqnarray}

Substituting Eqs.~(\ref{rel_mu_lambda_yneq0})
and (\ref{rel_x_y_yneq0}) into Eq.~(\ref{x_evol_eq_exp}),
leads to the following two constant solutions for $x$: 
\begin{eqnarray}
x= \frac{\sqrt{\tilde{\gamma}}\lambda}
{\sqrt{3(1+\tilde{\gamma})}},\;\;\;\;
{\rm or }\;\;\;\;x=\frac{\sqrt{3}(1+w_m)}
{\sqrt{ \tilde{\gamma} (1+\tilde{\gamma})} \lambda}.
\label{cond_kp_kps}
\end{eqnarray}
The first of these  solutions leads to  
$x^2 + y^2 =1$, which from Eqs.~(\ref{constraint}) and (\ref{omega_phi}) implies that 
during scaling $\Omega_m=0$ and $\Omega_{\phi}=1$. It 
corresponds to the relativistic kinetic-potential scaling solutions
given by
\begin{eqnarray}
(c4)&&(x,y,\tilde{\gamma})=
\biggl( \frac{\sqrt{\tilde{\gamma}}\lambda}
{\sqrt{3(1+\tilde{\gamma})}},
\sqrt{1-\frac{\tilde{\gamma} \lambda^2}{3 (1+\tilde{\gamma})}}
,\tilde{\gamma}\biggr),\nonumber\\
&& 
{\rm with} \;\;0<\tilde{\gamma}<1,\;\;
\lambda < 
\sqrt{\frac{3(1 + \tilde{\gamma})}{\tilde{\gamma}}},
\;\;\mu=-\lambda.\nonumber\\
\label{fixed_point_kinetic_potential}
\end{eqnarray} 
From Eqn.~(\ref{w_phi}) it implies $w_\phi = \frac{\tilde{\gamma} \lambda^2}{3}-1$ which means that the solution is accelerating if $\lambda^2 \tilde{\gamma} < 2$. Since for this solution, the DBI field completely dominates the fluid, this solution exists irrespective of the value of $w_m$.

Although somewhat complicated, the constraint
Eq.~(\ref{const_gamma_ito_xy_ex_sp}),
allows us to determine the actual values for $\tilde{\gamma}$, $x$ and $y$
for this fixed-point solution in terms of the parameters $\lambda$, $\sigma$, $\nu$.
For example, for $\sigma \nu >2/3$ they are given by
\begin{eqnarray}
\tilde{\gamma} &=& \frac{6}{\lambda^2 + 
\sqrt{36 + 12 (-1 + 3 \sigma \nu) \lambda^2 + \lambda^4}}\,,
\label{c4_gamma}
\\
x^2 &=&\frac{2 \lambda^2}{6+\lambda^2 + 
\sqrt{36 + 12 (-1 + 3 \sigma \nu) \lambda^2 + \lambda^4}}\,,
\label{c4_x}\\
y^2 &=& 1-x^2\,,
\label{c4_y}
\end{eqnarray} 
while for $\sigma \nu  < 2/3$, 
a related but slightly different expression holds. 

The second solution in Eq.~(\ref{cond_kp_kps}) 
corresponds to the relativistic kinetic-potential-fluid
scaling solution and is characterized by
\begin{eqnarray}
(c5)&&(x,y,\tilde{\gamma})\nonumber\\
&&=
\biggl(\frac{\sqrt{3}(1+w_m)}
{\sqrt{ \tilde{\gamma} (1+\tilde{\gamma})} \lambda},
\sqrt{\frac{3(1+w_m)(\tilde{\gamma} -w_m)}
{\lambda^2 \tilde{\gamma} (1+\tilde{\gamma})}},
\tilde{\gamma}
\biggr),\nonumber\\
&& 
{\rm with} \;\;0<\tilde{\gamma}<1,\;\; 
w_m < 0, \nonumber\\
&&\;\;\;\;\;\;\;\;\;\;\;\;
\lambda \geq \sqrt{\frac{3(1+w_m)}{\tilde{\gamma}}},
\;\;\mu = -\lambda.
\label{fixed_point_kinetic_potential_fluid}
\end{eqnarray} 

From Eqns.~(\ref{omega_phi}) and (\ref{w_phi}) we see that it leads to
$\Omega_\phi = \frac{3(1+w_m)}{\lambda^2 \tilde{\gamma}}$ and $w_\phi =
w_m$. In other words the DBI field equation of state tracks that of the
background matter. Since the property of this solution
is completely affected by the nature of the fluid,
the existence condition of this solution depends on
the value of $w_m$. It is worth noting that 
when we take into account the constraint Eq.~(\ref{const_gamma_ito_xy_ex_sp}), then from 
Eq.~(\ref{w_phi_sp_exp}), $w_\phi <0$.
Therefore, although it cannot be seen from 
Eq.~(\ref{fixed_point_kinetic_potential_fluid})
explicitly, this solution exists only for 
the case with $w_m <0$.

As in the case of $(c4)$, we can solve explicitly for $\tilde{\gamma}$,  $x$ and $y$
in  terms of $\lambda$, $\sigma$, $\nu$, $w_m$
by making use of the constraint given by Eq.~(\ref{const_gamma_ito_xy_ex_sp}).
For example, for the case with $\sigma \nu >1/3$, 
\begin{eqnarray}
\tilde{\gamma} &=& \frac{-2 w_m}{(3 \sigma \nu-1) (1+w_m)}
\nonumber\\
&&\times \left[\frac{1}{1-\sqrt{\frac{-4 w_m}
{(1-3 \sigma \nu)^2 (1+w_m)^2}+1}}\right]\,,
\label{c5_gamma}
\\
y^2 &=&\frac{18 \sigma \nu (1+w_m)}{(3 \sigma \nu-1)
(1+w_m) \lambda^2}\nonumber\\
&&\times \frac{1}{\left[2+(3 \sigma \nu -1)
\left(-1 +\sqrt{1 + \frac{1}{(3 \sigma \nu-1)^2 (1+w_m)^2}}
\right)\right]}\,,\nonumber\\
\label{c5_y}\\
&&x^2 = \frac{1+w_m}{\tilde{\gamma} -w_m} y^2\,,
\label{c5_x}
\end{eqnarray} 
while for $\sigma \nu  < 2/3$, 
a slightly different expression holds.

In summary we see that
there are ten fixed-point solutions for the case
where both the potential and brane tension are 
exponential functions of
the DBI field. In TABLE~\ref{fixed-points-Summary_exp} we summarise the  
two solutions $(c4)$ and $(c5)$ which appear when the particular condition $\lambda = -\mu$ is satisfied. The other solutions correspond to the case $q=0$ and are given in TABLE~\ref{fixed-points-Summary_power}.

\begin{widetext}
\begin{center}
\begin{table} [h!]
\begin{tabular} { |c|c|c|c|c|c|c| }
\hline 
 & $x$ & $y$ & $\tilde{\gamma}$ & $\Omega_\phi$ &  
Existence & Stability \\
\hline
$(c4)$ & $ \frac{\sqrt{\tilde{\gamma}}\lambda}
{\sqrt{3(1+\tilde{\gamma})}}$ 
& $\sqrt{1-\frac{\tilde{\gamma} \lambda^2}{3 (1+\tilde{\gamma})}}$ & $\tilde{\gamma}$ 
& $1$ 
&  
$0<\tilde{\gamma}<1, \lambda < 
\sqrt{\frac{3(1 + \tilde{\gamma})}{\tilde{\gamma}}},
\mu=-\lambda, \forall w_m$ & $\lambda < \sqrt{\frac{3(1+w_m)}{\tilde{\gamma}}}$ \\      
\hline
$(c5)$ & $ \frac{\sqrt{3}(1+w_m)}
{\sqrt{ \tilde{\gamma} (1+\tilde{\gamma})} \lambda}
$ 
& $\sqrt{\frac{3(1+w_m)(\tilde{\gamma} -w_m)}
{\lambda^2 \tilde{\gamma} (1+\tilde{\gamma})}}
$& $\tilde{\gamma}$ 
& $\frac{3(1+w_m)}{\tilde{\gamma} \lambda^2}$ 
& 
$0<\tilde{\gamma}<1, w_m < 0,
\lambda \geq \sqrt{\frac{3(1+w_m)}{\tilde{\gamma}}},
\mu=-\lambda$ & $\lambda \left(\geq \sqrt{\frac{3(1+w_m)}{\tilde{\gamma}}}
\right), w_m (< 0) $\\      
\hline
\end{tabular}
\caption{Summary of the extra 
fixed-points and their stability in the models 
where the potential and brane tension are exponential 
functions of the DBI field satisfying $\lambda = -\mu$. 
It can be shown that for $(c5)$, if the existence conditions
are satisfied, then the stability conditions are automatically
satisfied. Although we keep $\tilde{\gamma}$,
from the constraint given by Eq.~(\ref{const_gamma_ito_xy_ex_sp}), it is
 specified in terms of $\lambda$, $\sigma$, $\nu$, $w_m$
through Eq.~(\ref{c4_gamma}) for $(c4)$ and 
Eq.~(\ref{c5_gamma}) for $(c5)$.
In addition to these two, the solutions 
$(a1),\,(a2),\,(b1)-(b3),\,(c1),\,(c2), (\alpha 2)$ 
obtained in Sec.~\ref{subsec_pow_fixed} are also fixed-point 
solutions with the replacement 
$\tilde{\lambda} (= |p|) \to \lambda$
and $\tilde{\mu} (=r |p|/p) \to \mu$.
Notice that we have restricted ourselves to $\lambda >0$ and $x \geq 0$.}
\label{fixed-points-Summary_exp}
\end{table}
\end{center}
\end{widetext}

\subsection{Stability Analysis}
\label{stability-exp-pot}

We turn now to study the stability of the fixed-point solutions obtained 
in the previous subsection. Given that eight
of the fixed-point solutions
($(a1),\,(a2),\,(b1)-(b3),\,(c1),\,(c2), (\alpha 2)$) 
were also obtained and their stability determined
in Sec.~\ref{section_powerlaw} 
we do not repeat that analysis here. 
It is basicially the same as in that case except we make the  
replacement   
$\tilde{\lambda} (= |p|) \to \lambda$
and $\tilde{\mu} (=r |p|/p) \to \mu$.
Here, we concentrate on the two extra solutions $(c4)$ and 
$(c5)$ in Eqs.~(\ref{fixed_point_kinetic_potential}) 
and (\ref{fixed_point_kinetic_potential_fluid}). 
For the fixed-point $(c4)$ 
(relativistic kinetic-potential scaling solutions), having perturbed about the solution to linear order, we obtain for the corresponding eigenvalues:
\begin{eqnarray}
&&w_1 = \frac{\lambda^2 \tilde{\gamma}-6}{2} \nonumber\\
&& w_2 = \lambda^2 \tilde{\gamma}-3(1+w_m)\nonumber\\
&&w_3 = 0\,.
\end{eqnarray}
$w_3$ is zero because this corresponds to the $\tilde{\gamma}$
direction and at this fixed point, the RHS of 
Eq.~(\ref{gamma_evol_eq_exp}) becomes 0. 
As mentioned earlier, this is to be expected because 
there is a constraint on $\tilde{\gamma}$ and 
 only two of the three degrees of freedom 
are actually independent for $\lambda = - \mu$. Therefore  the existence of 
two negative eigenvlues will guarantee the stability
of this fixed-point solution.

Since $\lambda < \sqrt{3 (1+\tilde{\gamma}) /\tilde{\gamma}} 
< \sqrt{6 \tilde{\gamma}}$, $w_1$ is always negative.
Therefore,  solution $(c5)$  
is stable if $w_2 < 0$, that is, 
$\lambda < \sqrt{3(1+w_m)/\tilde{\gamma}}$. 
Recall that the condition
for inflation in this case is $\lambda^2 \tilde{\gamma} <2$. 

It is worth mentioning that by adopting the constraint
given by Eq.~(\ref{const_gamma_ito_xy_ex_sp}),
actually $\tilde{\gamma}$ is completely specified
in terms of $\lambda$, $\sigma$, $\nu$.
For example if $\sigma \nu > 2/3$, it can be shown that
this solution is stable if 
\begin{eqnarray} 
&&[6(3 + (3 \sigma \nu -1) \lambda^2)]w_m 
+18 - 6(1-3 \sigma \nu) \lambda^2+\lambda^4 \nonumber\\
&&\;\;\;\;\;\;\;\;
>
\sqrt{36-12(1-3 \sigma \nu) \lambda^2 + \lambda^4}\lambda^2\,.
\nonumber\\
\label{stab_cond_c4}
\end{eqnarray}

Similarly, for the fixed-point $(c5)$  
(relativistic kinetic-potential-fluid scaling solutions) we obtain:
\begin{eqnarray}
&&w_{1,2} = \frac{3}{4}(1-w_m)\nonumber\\
&&\times
\biggl[-1 \pm \sqrt{1-\frac{8 (1+w_m)(w_m-\tilde{\gamma}^2)
(3+3 w_m -\lambda^2 \tilde{\gamma})}
{\lambda^2 (-1+w_m)^2 \tilde{\gamma}}}\biggr]
\nonumber\\
&&w_3 = 0\,.
\end{eqnarray}

The same reasoning as applied to $(c4)$ suggests that even 
though $w_3=0$, the solution $(c5)$ is stable if the real parts of 
both $w_1$ and $w_2$ are negative. This requires 
$w_m  < \tilde{\gamma}^2$, which is automatically
satisfied if the existence condition $w_m < 0$
is satisfied,  although it would imply an unusual form of background matter being considered.

As in the case of $(c4)$, by adopting the constraint
given by Eq.~(\ref{const_gamma_ito_xy_ex_sp}),
actually $\tilde{\gamma}$ is completely specified
in terms of $\lambda$, $\sigma$, $\nu$, $w_m$.
It can be shown that if $\sigma \nu > 1/3$,
the existence condition 
$\lambda \geq \sqrt{3(1+w_m)}/\sqrt{\tilde{\gamma}}$
which is the most stringent
condition for this solution to be stable
can be expressed as
\begin{eqnarray} 
&&[6(3 + (3 \sigma \nu -1) \lambda^2)]w_m 
+18 - 6(1-3 \sigma \nu) \lambda^2+\lambda^4 \nonumber\\
&&\;\;\;\;\;\;\;\;
\leq
\sqrt{36-12(1-3 \sigma \nu) \lambda^2 + \lambda^4}\lambda^2\,.
\nonumber\\
\label{stab_cond_c5}
\end{eqnarray}

Clearly, Eqs.~(\ref{stab_cond_c4}) and (\ref{stab_cond_c5})
show that for $\sigma \nu >2/3$, there is no overlap
in the parameter space between where $(c4)$  and $(c5)$ are the late-time attractors. 
In TABLE~\ref{fixed-points-Summary_exp}, 
we summarise the stability conditions of 
these two fixed-point solutions.

 There has been related work  looking at the fixed-point
solutions $(c4)$ and $(c5)$. Actually, 
$(c4)$ was previously obtained in
\cite{Ahn:2009hu,Ahn:2009xd}, while
the existence of the scaling solution ($c5$) was
pointed out by Martin and Yamaguchi in \cite{Martin:2008xw}. 
The authors' stability analysis followed that of 
\cite{Ratra_Peebles}, which although providing a proof that 
the fixed-point solution is attractive, it did not
explain how the the solution could be realised from all initial values.
In our approach we have gone into more details, following the analysis of 
\cite{Copeland:1997et}. In particular by  making the phase space compact,
we have been able to establish all the fixed-points in the compact phase space. As we will now see it then becomes possible to discuss the late-time attractor
structure, something we now turn our attention to in the following subsection. The solution  $(c5)$
has recently led to a number of papers investigating its cosmology 
\cite{Martin:2008xw,Ahn:2009hu,Chiba:2009nh,Ahn:2009xd}

\subsection{Late-time behaviour}
\label{late-time-exp-pot}

In Sec.~\ref{section_powerlaw},
Fig.~\ref{fig_zero_attractor} shows the late-time attractor
structure with $q=0$ and $w_m >0$.
From the discussion at the beginning of this section,
except the special case with $\mu = -\lambda$,
Fig.~\ref{fig_zero_attractor} shows also  
the late-time attractor structure in the models 
where the potential and brane tension are exponential functions
of the DBI field with the replacement 
$\tilde{\lambda} (= |p|) \to \lambda$
and $\tilde{\mu} (=r |p|/p) \to \mu$. In the case with $\mu >-\lambda$, 
for $\lambda < \sqrt{3(1+w_m)}$,
the fixed-point $(b2)$ 
(standard kinetic-potential scaling solutions) 
is the late-time attractor, 
whereas for $\lambda \geq \sqrt{3(1+w_m)}$,
the fixed point $(b3)$ (standard kinetic-potential-fluid scaling solutions)
is a late-time attractor. For $\mu <- \sqrt{3(1+w_m)}/\sqrt{w_m}$
with $w_m >0$, in addition to $(b3)$, there is a possibility that  the fixed-point $(c1)$ 
(relativistic kinetic dominated solutions)
is also a late-time attractor. 
Because, they are all stable locally, which of these solutions actually wins out depends
on the initial value of $x$, $y$ and $\tilde{\gamma}$. 

On the other hand, from the discussions
in Sec.~\ref{section_powerlaw}, when $\mu < -\lambda$, 
the fixed-point solution $(\alpha 2)$ 
(ultra-relativistic potential dominated one) is the late-time
attractor. 
For the case  $\mu=-\lambda$, as  we have just seen, the additional fixed-points solutions 
$(c4)$ and
$(c5)$ are also possible late-time attractors, although the conditions under which they become attractors do not overlap in the available parameter space, so they do not compete with one another for overall stability. 

Of course when considering the overall stability for the  case of $\mu=-\lambda$, we need to 
also include a discussion of the fixed-point solutions 
which can be late-time attractors for the general case $\mu \neq \lambda$.
Since $(c1)$ and $(\alpha 2)$ do not satisfy
the constraint given by Eq.~(\ref{const_gamma_ito_xy_ex_sp}),
these two fixed-points can not be the late-time attractor
for $\mu = -\lambda$. It can be shown that 
$(c4)$ and $(c5)$ reduce to $(b2)$ and $(b3)$,
in the limit of $\tilde{\gamma} \to 1$. 

It follows that only $(c4)$ and $(c5)$ are the late-time attractors
for $\mu = -\lambda$ and once we specify the values
of the parameters,  $\lambda$, $\sigma$, $\nu$ and 
$w_m$ we can judge which of these two fixed-points will be
the late-time attractor. 

\subsection{Power-law models with $p+r=0$}
\label{subsec_special_powerlaw}

Here, for completeness,
it is appropriate to mention what happens when 
the models $V(\phi)$ and $f(\phi)$ are given by
\begin{eqnarray}
V(\phi) = \sigma |\phi|^p\,,\;\;\;\;
f(\phi) = \nu |\phi|^{-p}\,. 
\label{sp_powerlaw_model}
\end{eqnarray}

From (\ref{sp_powerlaw_model}), as in the exponential potential case with  $\mu = -\lambda$ discussed in section~\ref{fixed-point-exp-pot}, there is a constraint
given by Eq.~(\ref{const_gamma_ito_xy_ex_sp}),
implying that only two of the three variables are independent.
This in turn implies that we cannot make use of the
degree of freedom corresponding to $f(\phi)$
to construct an autonomous system to solve 
Eqs.~(\ref{x_evol_eq_exp}) - (\ref{mu_evol_eq}) as they stand.
So if there is a constraint of the form $f(\phi) V(\phi)=const.$,
the only case we can obtain an autonomous system
is when $V \sim \exp[-\lambda \phi]$ and 
$f \sim \exp [\lambda \phi]$ (or equivalently $p \to \infty$).

We can of course still make some progress. For example for the case of a canonical scalar field, not described as an autonomous system, the late-time behaviour
with a power-law potential has been determined in 
\cite{delaMacorra:1999ff,Ng:2001hs,Mizuno:2004xj}.
We adopt a similar procedure here for the case described by  Eq.~(\ref{sp_powerlaw_model}), by 
regarding these cases as 
limits of the models where the potential and 
brane tension are exponential functions of the DBI
field satisfying $\mu=-\lambda$. The point is
that in terms of 
$\lambda$ and $\mu$, 
the evolution equations for $x$, $y$, $\tilde{\gamma}$  
are still given by
Eqs.~(\ref{x_evol_eq_exp}), (\ref{y_evol_eq_exp}) and 
(\ref{gamma_evol_eq_exp}). The difference from the cases
with the exponential potential and brane tension is
that $\lambda$ and $\mu$  are not constant for the power-law cases
but are given by 
\begin{eqnarray} 
\lambda=-\mu = - \varepsilon p \frac{1}{|\phi|}.  
\end{eqnarray}

As in the other cases, we can generally restrict
$\lambda > 0$ which is equivalent to considering
only $\phi <0$ for $p>0$ and $\phi >0$ for $p<0$.
Therefore, in the late-time limit, $\phi \to 0$ for $p > 0$
while $\phi \to \infty$ for $p < 0$.
This means that the late-time asymptotic value 
of $\lambda$ is $\lambda \to + \infty$ for $p >0$,
while $\lambda \to + 0$ for $p < 0$.

First, let us consider the case 
with $p < 0$ ($\lambda \to 0$). In this case,
from Table~\ref{fixed-points-Summary_exp} 
and Fig.~\ref{fig_zero_attractor}, the only possible
late-time attractor solution is 
$(c4)$ (relativistic kinetic-potential
scaling solutions). In the limit $\lambda \to 0$,
from Eqs.(\ref{c4_gamma}) - (\ref{c4_y}),
it asymptotes to $(x,y,\tilde{\gamma})=(0,1,1)$,
that is, standard potential dominated solutions.
Therefore, 
the DBI-field behaves like a cosmological constant
at late-time.

Next, let us consider the cases with $p>0$ 
($\lambda \to \infty$ ). In these cases, since $\lambda$
is very large, from 
Table~\ref{fixed-points-Summary_exp} 
and Fig.~\ref{fig_zero_attractor}, 
the only possible
late-time attractor solution is 
$(c5)$ (relativistic kinetic-potential-fluid
scaling solutions).

In the limit $\lambda \to \infty$,
from Eqs.(\ref{c5_gamma}) - (\ref{c5_x}),
it asymptotes to $(x,y,\tilde{\gamma})=(0,0,\tilde{\gamma})$,
that is fluid dominated solution and
the DBI field becomes irrelevant compared to the fluid.
It is worth noting that from Eq.~(\ref{c5_gamma})
$\tilde{\gamma}$ is completely specified
once $\sigma$, $\nu$, $w_m$ are fixed. 

\section{Summary\label{sec_summary}}

Successful models of inflation arising within string theory from the DBI action have generated a great deal of interest recently. With their non-canonical kinetic terms, non-trivial potentials and brane tensions, they have led to a number of fascinating results including the prediction of distinctive non-Gaussian fluctuations in the CMB. Although most models investigated to date have had specific functional forms  for the potential and brane tension, in this paper we have decided to broaden the class of models being discussed and so have analysed the dynamics associated with more general forms for these potential and brane tension functions, including in the analysis the presence of a background perfect fluid in a flat FRW universe. Following the approach developed in \cite{Copeland:1997et}, we have introduced a suitable set of dynamical variables $x$, $y$ and $\tilde{\gamma}$ in Eq.~(\ref{defn_xy}) which has allowed us to determine the phase-space portrait of the system. In particular, we have established the late time behaviour of these systems, demonstrating where appropriate the attractor nature of the solutions. 

In Sec.~\ref{section_powerlaw}, we have considered the models
where the potential and brane tension are given by 
power-law functions of the DBI field
($V(\phi) = \sigma |\phi|^p$, $f(\phi) = \nu |\phi|^r$).
The standard fixed-point solutions of this system 
are summarised
in TABLE~\ref{fixed-points-Summary_power}, where we see that the late-time attractor
nature of the solutions depends on $q \equiv -1/(p+r)$. The interesting cases of scaling where the ratio of the kinetic to potential energies of the DBI field is a constant are found to exist only for  $q=1/2$ ($(a4)$, $(a5)$) and  $q=0$ ($(b2)$, $(b3)$).
This is because if we require $x$ and $y$ to be nonzero constants,
there are only two possibilities, that is, $\tilde{\gamma}=0$ (leading to scaling with $q=1/2$)  
and  $\tilde{\gamma}=1$ (leading to scaling with $q=0$). 
These scaling solutions are then shown to be stable for certain regions
of the parameter space. In addition to these, we have also
explicitly demonstrated the existence and stability of an
interesting inflationary solution $(c3)$ 
specific to the DBI field with constant
$\tilde{\gamma}$ which differs from $\tilde{\gamma}=1$ and $\tilde{\gamma}=0$. 

The DBI system is rich. For example the evolution equations (\ref{x_evol_eq_mod})-(\ref{gamma_evol_eq_mod}) can appear singular when some of $x$, $y$ or $\tilde{\gamma}$ either tend to zero or unity, which one is singular depends on the value of $q$. On the face of it, the equations appear to be ill-defined, but in practice it turns out that these points can actually be  late-time attractor solutions for the system. In section~\ref{fixedpoints0/0} we obtain these fixed points and determine their stability. These are summarised in Table~\ref{fixed-points-Summary_0/0}.
 
Having established all the fixed-points and their stability for a given set of parameters, 
we have gone on to determine which of these solutions
will be the late-time attractor. Particular care is required when considering the cases 
where the eigenvalues associated with the perturbations vanish, implying that the solution is 
marginally stable. Whether these are the late time attractors for the system 
depends upon the stability
of the other allowed fixed-points. In particular 
those fixed points with three negative eigenvalues,
means that these marginally stable fixed-point solutions
can not be the late-time attractor as their stability
is weaker than that of the fixed-points. However,
if all other fixed-points have positive eigenvalues,
then the marginally stable fixed-point solution can
turn out to be the late-time attractor.
We have summarised the possible late-time attractor solutions
for a given set of $q$, $p$, $r$ in Table~III.

In Sec.~\ref{section_exp}, we have considered the models
where the potential and brane tension are 
exponential functions of the DBI field
($V(\phi) = C e^{-\lambda \phi}$, $f(\phi) = D e^{-\mu \phi})$.
This system has a similar dynamical structure to that of the 
power-law models with $q=0$. However, there are additional scaling solutions present in the exponential case because we can construct an autonomous system for the case  with $\lambda = -\mu$, leading to solutions where $\tilde{\gamma}$ is a constant between $0$ and $1$
($(c4)$,$(c5)$). In this case, we find that once we specify
the values of the parameters, $\lambda$, $\sigma$, $\nu$
and $w_m$ we can judge which of these two fixed-points
will be the late-time attractor. The stability of these
two fixed-points are summarised in Table~IV.  We have also shown that the special case 
($V(\phi) = \sigma |\phi|^p$, $f(\phi) = \nu |\phi|^{-p}$)
can be discussed in terms of the limiting behaviour $\lambda \to 0$ or 
$\lambda \to \infty$ in the models with an exponential 
potential and brane tension satisfying $\lambda=-\mu$. 

There is an overlap between elements of this work and other published material. For
example,  we are able to reproduce a number of results obtained earlier
in  \cite{Guo:2008sz} where 
the authors considered the case with a massive potential and AdS throat.
It turns out to be a special case of $q=-1/2$ discussed in Sec.~III in this paper.

The existence of the scaling solution ($c5$) was originally 
pointed out  by Martin and Yamaguchi in \cite{Martin:2008xw}. 
The authors stability analysis followed that of 
\cite{Ratra_Peebles}, which although providing a proof that 
the fixed-point solution is attractive, it did not
explain how the the solution could be realised from all initial values.
In our approach we have gone into more details, following the analysis of 
\cite{Copeland:1997et}. In particular by  making the phase space compact,
we have been able to establish all the fixed-points in the compact phase
space, allowing us to then properly discuss the late-time attractor
structure of the solution.

Of most interest to us though is the existence 
and stability of a  class of cosmologically relevant 
solutions. We have found that a fixed point solution $(c3)$ 
(relativistic potential dominated solutions)
where $\tilde{\gamma}$  is a constant satisfying $ 0 < \sqrt{3}/\sqrt{\tilde{\lambda}^2 + 3} < 1$ 
can be the late-time attractor for $q=-1/2$ with $r/p < -1$
(see also \cite{Ahn:2009xd}).
Although it can not be applied to the AdS throat ($r=-4$)
as $r/p > -1$, we believe this solution is very interesting and important.
In calculations of primordial perturbations
of the DBI inflation models, 
properly incorporating the time dependence of
the sound speed $c_s$ is a complicated issue 
\cite{Garriga:1999vw,Alishahiha:2004eh}, and in fact it is usually assumed to be constant.
(For a recent approach to relaxing this assumption, see \cite{Lorenz:2008et}.)
Since in our case $c_s=\tilde{\gamma}$  
is constant for the fixed-point solution $(c3)$, 
this serves as a good background to use when we 
consider cosmological perturbations in DBI inflation.


\section*{Acknowledgments}
We would like to thank Kazuya Koyama, Jerome Martin,
Shinji Mukohyama, Ryo Saito, David Wands, Masahide Yamaguchi 
and Jun'ichi Yokoyama for interesting discussions.
We also would like to thank Eric Linder for 
deteiled discussions concerning some of the solutions.  
S. M. is grateful to the RESCEU, the University of Tokyo
for their hospitality when this work was
completed.
S. M. is supported by JSPS Postdoctral Fellowships 
for Research Abroad.
E. J. C. is grateful to the Royal Society for financial support.



\begin{references}

\bibitem{Linde}
A.~D.~Linde,
{\em Particle Physics and Inflationary  Cosmology},
(Harwood academic publishers, 1980).

\bibitem{Lyth:1998xn}
  D.~H.~Lyth and A.~Riotto,
  Phys.\ Rept.\  {\bf 314}, 1 (1999)
  [arXiv:hep-ph/9807278].

\bibitem{Baumann:2009ni}
  D.~Baumann and L.~McAllister,
  Ann.\ Rev.\ Nucl.\ Part.\ Sci.\  {\bf 59} (2009) 67
  [arXiv:0901.0265 [hep-th]].


\bibitem{Silverstein:2003hf}
  E.~Silverstein and D.~Tong,
  Phys.\ Rev.\  D {\bf 70} (2004) 103505
  [arXiv:hep-th/0310221].

\bibitem{Alishahiha:2004eh}
  M.~Alishahiha, E.~Silverstein and D.~Tong,
  Phys.\ Rev.\  D {\bf 70} (2004) 123505
  [arXiv:hep-th/0404084].

\bibitem{Chen:2004gc}
  X.~Chen,
  Phys.\ Rev.\  D {\bf 71} (2005) 063506
  [arXiv:hep-th/0408084].


\bibitem{Chen:2005ad}
  X.~Chen,
  JHEP {\bf 0508} (2005) 045
  [arXiv:hep-th/0501184].

\bibitem{Shandera:2006ax}
  S.~E.~Shandera and S.~H.~Tye,
  JCAP {\bf 0605} (2006) 007
  [arXiv:hep-th/0601099].






\bibitem{Spalinski:2007dv}
  M.~Spalinski,
  JCAP {\bf 0705}, 017 (2007)
  [arXiv:hep-th/0702196].


\bibitem{Spalinski:2007qy}
  M.~Spalinski,
  Phys.\ Lett.\  B {\bf 650} (2007) 313
  [arXiv:hep-th/0703248].

\bibitem{Chimento:2007es}
  L.~P.~Chimento and R.~Lazkoz,
  Gen.\ Rel.\ Grav.\  {\bf 40}, 2543 (2008)
  [arXiv:0711.0712 [hep-th]].


\bibitem{Ward:2007gs}
  J.~Ward,
  JHEP {\bf 0712}, 045 (2007)
  [arXiv:0711.0760 [hep-th]].

\bibitem{Spalinski:2007un}
  M.~Spalinski,
  JCAP {\bf 0804} (2008) 002
  [arXiv:0711.4326 [astro-ph]].

\bibitem{Kinney:2007ag}
  W.~H.~Kinney and K.~Tzirakis,
  Phys.\ Rev.\  D {\bf 77} (2008) 103517
  [arXiv:0712.2043 [astro-ph]].



\bibitem{Tzirakis:2008qy}
  K.~Tzirakis and W.~H.~Kinney,
  JCAP {\bf 0901} (2009) 028
  [arXiv:0810.0270 [astro-ph]].

\bibitem{Czuchry:2008km}
  E.~Czuchry,
  Phys.\ Lett.\  B {\bf 678} (2009) 9
  [arXiv:0812.1409 [astro-ph]].

\bibitem{Chen:2005fe}
  X.~Chen,
  Phys.\ Rev.\  D {\bf 72} (2005) 123518
  [arXiv:astro-ph/0507053].


\bibitem{Chen:2006nt}
  X.~Chen, M.~x.~Huang, S.~Kachru and G.~Shiu,
  JCAP {\bf 0701} (2007) 002
  [arXiv:hep-th/0605045].


\bibitem{Huang:2006eh}
 X.~Chen, M.~x.~Huang and G.~Shiu,
  Phys.\ Rev.\  D {\bf 74} (2006) 121301
  [arXiv:hep-th/0610235].

\bibitem{Arroja:2008ga}
  F.~Arroja and K.~Koyama,
  Phys.\ Rev.\  D {\bf 77}, 083517 (2008)
  [arXiv:0802.1167 [hep-th]].

\bibitem{Langlois:2008wt}
  D.~Langlois, S.~Renaux-Petel, D.~A.~Steer and T.~Tanaka,
  Phys.\ Rev.\ Lett.\  {\bf 101}, 061301 (2008)
  [arXiv:0804.3139 [hep-th]].

\bibitem{Langlois:2008qf}
  D.~Langlois, S.~Renaux-Petel, D.~A.~Steer and T.~Tanaka,
  Phys.\ Rev.\  D {\bf 78}, 063523 (2008)
  [arXiv:0806.0336 [hep-th]].


\bibitem{Arroja:2008yy}
  F.~Arroja, S.~Mizuno and K.~Koyama,
  JCAP {\bf 0808}, 015 (2008)
  [arXiv:0806.0619 [astro-ph]].

\bibitem{Langlois:2009ej}
  D.~Langlois, S.~Renaux-Petel and D.~A.~Steer,
  JCAP {\bf 0904} (2009) 021
  [arXiv:0902.2941 [hep-th]].

\bibitem{Gao:2009gd}
  X.~Gao and B.~Hu,
  JCAP {\bf 0908} (2009) 012
  [arXiv:0903.1920 [astro-ph.CO]].

\bibitem{Chen:2009bc}
  X.~Chen, B.~Hu, M.~x.~Huang, G.~Shiu and Y.~Wang,
  JCAP {\bf 0908}, 008 (2009)
  [arXiv:0905.3494 [astro-ph.CO]].


\bibitem{Us}
 F.~Arroja, S.~Mizuno, K.~Koyama and T.~Tanaka,
  Phys.\ Rev.\  D {\bf 80} (2009) 043527
  [arXiv:0905.3641 [hep-th]].

\bibitem{Mizuno:2009cv}
  S.~Mizuno, F.~Arroja, K.~Koyama and T.~Tanaka,
  Phys.\ Rev.\  D {\bf 80}, 023530 (2009)
  [arXiv:0905.4557 [hep-th]].



\bibitem{Gao:2009at}
  X.~Gao, M.~Li and C.~Lin,
  JCAP {\bf 0911} (2009) 007
  [arXiv:0906.1345 [astro-ph.CO]].

\bibitem{Mizuno:2009mv}
  S.~Mizuno, F.~Arroja and K.~Koyama,
  Phys.\ Rev.\  D {\bf 80}, 083517 (2009)
  [arXiv:0907.2439 [hep-th]].


\bibitem{RenauxPetel:2009sj}
  S.~Renaux-Petel,
  JCAP {\bf 0910} (2009) 012
  [arXiv:0907.2476 [hep-th]].

\bibitem{Chen:2009zp}
  X.~Chen and Y.~Wang,
  arXiv:0911.3380 [hep-th].

\bibitem{Koyama:2010xj}
  K.~Koyama,
  arXiv:1002.0600 [hep-th].


\bibitem{Chen:2010xk}
  X.~Chen,
  arXiv:1002.1416 [astro-ph.CO].

\bibitem{Kecskemeti:2006cg}
  S.~Kecskemeti, J.~Maiden, G.~Shiu and B.~Underwood,
  JHEP {\bf 0609}, 076 (2006)
  [arXiv:hep-th/0605189].

\bibitem{Lidsey:2006ia}
J.~E. Lidsey and D.~Seery,
\newblock Phys. Rev. {\bf D75}, 043505 (2007), astro-ph/0610398.

\bibitem{Baumann:2006cd}
  D.~Baumann and L.~McAllister,
  Phys.\ Rev.\  D {\bf 75}, 123508 (2007)
  [arXiv:hep-th/0610285].

\bibitem{Bean:2007hc}
  R.~Bean, S.~E.~Shandera, S.~H.~Henry Tye and J.~Xu,
  JCAP {\bf 0705}, 004 (2007)
  [arXiv:hep-th/0702107].

\bibitem{Lidsey:2007gq}
  J.~E.~Lidsey and I.~Huston,
  JCAP {\bf 0707}, 002 (2007)
  [arXiv:0705.0240 [hep-th]].

\bibitem{Peiris:2007gz}
H.~V. Peiris, D.~Baumann, B.~Friedman, and A.~Cooray,
\newblock Phys. Rev. {\bf D76}, 103517 (2007), 0706.1240.

\bibitem{Kobayashi:2007hm}
  T.~Kobayashi, S.~Mukohyama and S.~Kinoshita,
  JCAP {\bf 0801}, 028 (2008)
  [arXiv:0708.4285 [hep-th]].


\bibitem{Gmeiner:2007uw}
  F.~Gmeiner and C.~D.~White,
  JCAP {\bf 0802}, 012 (2008)
  [arXiv:0710.2009 [hep-th]].

\bibitem{Lorenz:2007ze}
L.~Lorenz, J.~Martin, and C.~Ringeval,
\newblock JCAP {\bf 0804}, 001 (2008), 0709.3758.

\bibitem{Bean:2007eh}
  R.~Bean, X.~Chen, H.~Peiris and J.~Xu,
  Phys.\ Rev.\  D {\bf 77}, 023527 (2008)
  [arXiv:0710.1812 [hep-th]].


\bibitem{Bird:2009pq}
  S.~Bird, H.~V.~Peiris and D.~Baumann,
  Phys.\ Rev.\  D {\bf 80} (2009) 023534
  [arXiv:0905.2412 [hep-th]].

\bibitem{Bessada:2009pe}
  D.~Bessada, W.~H.~Kinney and K.~Tzirakis,
  JCAP {\bf 0909}, 031 (2009)
  [arXiv:0907.1311 [gr-qc]].

\bibitem{Fuzfa:2005qn}
  A.~Fuzfa and J.~M.~Alimi,
  Phys.\ Rev.\  D {\bf 73} (2006) 023520
  [arXiv:gr-qc/0511090].

\bibitem{Fuzfa:2006pn}
  A.~Fuzfa and J.~M.~Alimi,
  Phys.\ Rev.\ Lett.\  {\bf 97} (2006) 061301
  [arXiv:astro-ph/0604517].




\bibitem{Martin:2008xw}
  J.~Martin and M.~Yamaguchi,
  Phys.\ Rev.\  D {\bf 77}, 123508 (2008)
  [arXiv:0801.3375 [hep-th]].

\bibitem{Ahn:2009hu}
  C.~Ahn, C.~Kim and E.~V.~Linder,
  Phys.\ Lett.\  B {\bf 684}, 181 (2010)
  [arXiv:0904.3328 [astro-ph.CO]].

\bibitem{Chiba:2009nh}
  T.~Chiba, S.~Dutta and R.~J.~Scherrer,
  Phys.\ Rev.\  D {\bf 80}, 043517 (2009)
  [arXiv:0906.0628 [astro-ph.CO]].


\bibitem{Ahn:2009xd}
  C.~Ahn, C.~Kim and E.~V.~Linder,
  Phys.\ Rev.\  D {\bf 80}, 123016 (2009)
  [arXiv:0909.2637 [astro-ph.CO]].

\bibitem{Lucchin:1984yf}
  F.~Lucchin and S.~Matarrese,
  Phys.\ Rev.\  D {\bf 32}, 1316 (1985).

\bibitem{Ratra_Peebles}
B.~Ratra and P.~J.~E.~Peebles,
Phys.\ Rev.\ D {\bf 37}, 3406 (1988).

\bibitem{trac} C. Wetterich, Nucl. Phys. 
\textbf{B302}, 668 (1988)

\bibitem{Ferreira:1997hj}
  P.~G.~Ferreira and M.~Joyce,
  Phys.\ Rev.\  D {\bf 58} (1998) 023503
  [arXiv:astro-ph/9711102].

\bibitem{Copeland:1997et}
  E.~J.~Copeland, A.~R.~Liddle and D.~Wands,
  Phys.\ Rev.\  D {\bf 57} (1998) 4686
  [arXiv:gr-qc/9711068].

\bibitem{vdHCW}
  R.~J.~van den Hoogen, A.~A.~Coley and D.~Wands,
  Class.\ Quant.\ Grav.\  {\bf 16}, 1843 (1999)
  [arXiv:gr-qc/9901014].

\bibitem{Heard:2002dr}
  I.~P.~C.~Heard and D.~Wands,
  Class.\ Quant.\ Grav.\  {\bf 19} (2002) 5435
  [arXiv:gr-qc/0206085].

\bibitem{Padmanabhan:2002cp}
  T.~Padmanabhan,
  Phys.\ Rev.\  D {\bf 66} (2002) 021301
  [arXiv:hep-th/0204150].

\bibitem{Tsujikawa:2004dp}
  S.~Tsujikawa and M.~Sami,
  Phys.\ Lett.\  B {\bf 603} (2004) 113
  [arXiv:hep-th/0409212].

\bibitem{Calcagni:2004wu}
  G.~Calcagni,
  Phys.\ Rev.\  D {\bf 71}, 023511 (2005)
  [arXiv:gr-qc/0410027].


\bibitem{Copeland:2004qe}
E.~J.~Copeland, S.~J.~Lee, J.~E.~Lidsey and S.~Mizuno,
Phys.\ Rev.\ D {\bf 71}, 023526 (2005)
[arXiv:astro-ph/0410110].

\bibitem{Copeland:2009be}
  E.~J.~Copeland, S.~Mizuno and M.~Shaeri,
  Phys.\ Rev.\  D {\bf 79} (2009) 103515
  [arXiv:0904.0877 [astro-ph.CO]].



\bibitem{Meng:2004ap}
  X.~H.~Meng and P.~Wang,
  Class.\ Quant.\ Grav.\  {\bf 21}, L101 (2004)
  [arXiv:astro-ph/0406476].

\bibitem{Underwood:2008dh}
  B.~Underwood,
  Phys.\ Rev.\  D {\bf 78}, 023509 (2008)
  [arXiv:0802.2117 [hep-th]].

\bibitem{Franche:2009gk}
  P.~Franche, R.~Gwyn, B.~Underwood and A.~Wissanji,
  arXiv:0912.1857 [hep-th].

\bibitem{Franche:2010yj}
  P.~Franche, R.~Gwyn, B.~Underwood and A.~Wissanji,
  arXiv:1002.2639 [hep-th].





\bibitem{Guo:2008sz}
  Z.~K.~Guo and N.~Ohta,
  JCAP {\bf 0804}, 035 (2008)
  [arXiv:0803.1013 [hep-th]].


\bibitem{Aguirregabiria:2004xd}
  J.~M.~Aguirregabiria and R.~Lazkoz,
  Phys.\ Rev.\  D {\bf 69}, 123502 (2004)
  [arXiv:hep-th/0402190].

\bibitem{Piazza:2004df}
  F.~Piazza and S.~Tsujikawa,
  JCAP {\bf 0407} (2004) 004
  [arXiv:hep-th/0405054].

\bibitem{Copeland:2004hq}
  E.~J.~Copeland, M.~R.~Garousi, M.~Sami and S.~Tsujikawa,
  Phys.\ Rev.\  D {\bf 71}, 043003 (2005)
  [arXiv:hep-th/0411192].

\bibitem{Gumjudpai:2006hg}
  B.~Gumjudpai, T.~Naskar and J.~Ward,
  JCAP {\bf 0611}, 006 (2006)
  [arXiv:hep-ph/0603210].


\bibitem{Tsujikawa:2006mw}
  S.~Tsujikawa,
  Phys.\ Rev.\  D {\bf 73} (2006) 103504
  [arXiv:hep-th/0601178].

\bibitem{Gong:2006sp}
  Y.~Gong, A.~Wang and Y.~Z.~Zhang,
  Phys.\ Lett.\  B {\bf 636}, 286 (2006)
  [arXiv:gr-qc/0603050].


\bibitem{Sen:2008yt}
  A.~A.~Sen and N.~C.~Devi,
  Phys.\ Lett.\  B {\bf 668}, 182 (2008)
  [arXiv:0804.2775 [astro-ph]].





\bibitem{Quiros:2009mz}
  I.~Quiros, T.~Gonzalez, D.~Gonzalez and Y.~Napoles,
  arXiv:0906.2617 [gr-qc].

\bibitem{Li:2010eu}
  J.~L.~Li and J.~P.~Wu,
  arXiv:1003.1870 [hep-th].


\bibitem{Gumjudpai:2009uy}
  B.~Gumjudpai and J.~Ward,
  arXiv:0904.0472 [astro-ph.CO].

\bibitem{Saridakis:2009uk}
  E.~N.~Saridakis and J.~Ward,
  Phys.\ Rev.\  D {\bf 80} (2009) 083003
  [arXiv:0906.5135 [hep-th]].




\bibitem{Pajer:2008uy}
  E.~Pajer,
  JCAP {\bf 0804}, 031 (2008)
  [arXiv:0802.2916 [hep-th]].

\bibitem{Komatsu:2010fb}
  E.~Komatsu {\it et al.},
  arXiv:1001.4538 [astro-ph.CO].

\bibitem{delaMacorra:1999ff}
  A.~de la Macorra and G.~Piccinelli,
  Phys.\ Rev.\  D {\bf 61} (2000) 123503
  [arXiv:hep-ph/9909459].

\bibitem{Ng:2001hs}
  S.~C.~C.~Ng, N.~J.~Nunes and F.~Rosati,
  Phys.\ Rev.\  D {\bf 64}, 083510 (2001)
  [arXiv:astro-ph/0107321].

\bibitem{Mizuno:2004xj}
  S.~Mizuno, S.~J.~Lee and E.~J.~Copeland,
  Phys.\ Rev.\  D {\bf 70} (2004) 043525
  [arXiv:astro-ph/0405490].

\bibitem{Garriga:1999vw}
  J.~Garriga and V.~F.~Mukhanov,
  Phys.\ Lett.\  B {\bf 458}, 219 (1999)
  [arXiv:hep-th/9904176].

\bibitem{Lorenz:2008et}
  L.~Lorenz, J.~Martin and C.~Ringeval,
  Phys.\ Rev.\  D {\bf 78}, 083513 (2008)
  [arXiv:0807.3037 [astro-ph]].

\end{references}
\end{document}